\documentclass[twocolumn, showpacs,preprintnumbers,amsmath,amssymb, longbibliography]{revtex4-1}
\usepackage{graphicx}
\usepackage{dcolumn}
\usepackage{bm}

\begin{document}

\title{Giant deviation of a relaxation time from generalized Newtonian theory in Discontinuous Shear Thickening suspensions}
%Giant anomaly in the magnitude of a relaxation time in discontinuous shear thickening suspensions} % Title
\author{Rijan Maharjan} % Author name
\author{Eric Brown}

\affiliation{Department of Mechanical Engineering and Materials Science, Yale University, New Haven, CT 06520}
\date{\today} % Date for the report

%\begin{multicols*}{2}
%\doublespacing
\begin{abstract}
We investigated the transient relaxation of a Discontinuous Shear Thickening (DST) suspension of cornstarch in water.  We performed 2 types of relaxation experiments starting from a  steady shear  in a parallel plate rheometer, followed by either stopping the top plate  rotation and measuring the transient torque relaxation, or removing the torque on the plate and measuring the transient rotation of the tool.  We found that at low effective weight fraction $\phi_{eff}<58.8\pm0.4\%$, the suspensions exhibited a relaxation behavior consistent with a generalized Newtonian fluid in which the relaxation is determined by the steady-state relationship between shear stress and shear rate.  However, for larger weight fraction $58.8\% < \phi_{eff} < 61.0\%$, near the liquid-solid transition $\phi_c=61.0\pm0.7\%$, we found relaxation behaviors qualitatively  and quantitatively different from the generalized Newtonian model.   The regime where the relaxation was inconsistent with the generalized Newtonian model was the same  where we found positive normal stress during relaxation, and in some cases we found an oscillatory response, suggestive of  a solid-like structure consisting of a system-spanning contact network of particles.  This regime also corresponds to  the same packing fraction range where we consistently found  discontinuous shear thickening in rate-controlled, steady-state measurements.  The relaxation time in this range scales with the inverse of the critical shear rate at the onset of shear thickening, which may correspond to a contact relaxation time for nearby particles in the structure to flow away from each other.   In this range the  relaxation time was the  same in both stress- and rate-controlled relaxation experiments, indicating the relaxation time is more intrinsic than an effective viscosity in this range, and is needed in addition to the steady-state viscosity function to describe transient flows.   The discrepancy between the measured  relaxation times and the generalized Newtonian prediction was found to be as large as 4 orders of magnitude, and extrapolations diverge in the limit as $\phi_{eff} \rightarrow \phi_c$ as the generalized Newtonian prediction approaches 0.   This quantitative discrepancy indicates the relaxation is not controlled by the dissipative terms in the constitutive relation.    At the highest weight fractions, the relaxation time scales were  measured to be on the order of $\sim 1$ s.   The fact that this timescale is resolvable by the naked eye may be important to understanding some of the  dynamic phenomenon  commonly observed in cornstarch and water suspensions.  We also showed that using the critical shear rate $\dot\gamma_c$ at the onset of shear thickening to characterize the effective weight fraction $\phi_{eff}$ can more precisely characterize material properties near  the critical point $\phi_c$,  allowing us to resolve this transition so close to $\phi_c$.  This conversion to $\phi_{eff}$ can also be used  to compare experiments done in other laboratories  or under different temperature and humidity conditions on a consistent $\phi_{eff}$ scale at our reference temperature  and humidity environment.
\end{abstract}
\maketitle
%\section{Introduction}
\label{sec:Intro}

% definition and broad problem  statement
Discontinuous Shear Thickening (DST) suspensions are known to exhibit a number of transient phenomena which are associated with the ability to form a temporary solid-like jammed state in response to shear.  Such phenomena include an impact response strong enough to be used in commercial impact protection devices \cite{LWW03, D3O},  a shear resistance strong  enough to break or jam industrial mixing equipment \cite{Ba89}, the ability of people to run on the  surface of cornstarch and water \cite{youtube_running, BJ14},  the formation of stable holes in the surface of a vertically vibrated layer of the fluid \cite{MDGRS04}, and oscillations in the velocity of a sphere sinking in the fluid \cite{KSLM11}.  There is also an unjamming process for each of these phenomena, in which the solid-like state relaxes back to a fluid-like state, which determines how long the phenomena last and may be relevant for cyclic motion.   Our goal is to characterize the transient relaxation component of the rheology of DST suspensions to better understand such phenomena.

%background definitions -  steady-state DST
Traditionally, shear thickening is defined by an increase in effective viscosity $\eta$ with increasing shear rate $\dot\gamma$ or shear stress $\tau$, where $\eta=\tau/\dot\gamma$ in a steady-state flow in a rheometer. In many concentrated suspensions such as cornstarch and water,  this effect can be so strong that the increase of $\eta$ or $\tau$ in $\dot\gamma$ can be 1-3 orders of magnitude, and can even appear to be discontinuous in $\dot\gamma$ at a critical shear rate $\dot\gamma_c$.  The steepness of $\tau(\dot\gamma)$ tends to increase with packing fraction up to the liquid-solid transition $\phi_c$,  above which shear thickening is no longer observed \cite{BJ09, BZFMBDJ11}, so shear thickening is most prominent just below $\phi_c$  (see \citep{Ba89, WB09, BJ14} for reviews on shear thickening).      

%relaxation context -  transient and dynamic phenomena and  problems with the generalized Newtonian models
A typical purpose of rheology measurements is to  obtain a constitutive relation for the shear stress which could then be inserted into a modified Navier-Stokes equation in place of the  usual shear stress term, and solved to describe flows under different conditions (i.e.~geometries, boundary conditions, and transients). However, transient  impact experiments have revealed a very different rheology than  the steady-state $\tau(\dot\gamma)$ from rheometer experiments described in the previous paragraph.  For example, under impact DST suspensions support stresses orders of magnitude larger than inferred from steady-state rheometer measurements \cite{WJ12, PJ14, MMASB17}.  This contradicts the simplest rheological model of a generalized Newtonian fluid -- meaning the same constitutive relation $\tau(\dot\gamma)$  that has a single value of shear stress $\tau$ at each shear rate $\dot\gamma$ applies to all flows.  Thus, any complete rheology that includes both the steady-state and transient behavior of DST suspensions requires more information than just  $\tau(\dot\gamma)$.

% 2nd example
Another unusual phenomena of DST fluids is the formation of stable holes in the surface of a vertically vibrated layer of the fluid \cite{MDGRS04}. It has been shown that these structures cannot be stable due to a rheology described by any generalized Newtonian function of the form $\tau(\dot\gamma)$ -- regardless of whether the function includes shear thickening. Instead the $\tau(\dot\gamma)$ must have a hysteresis such that there is a difference in stress on the up- and down-cycles of the vibration to overcome the gravitational and surface tension forces that are trying to close the hole \cite{De10}. 

%3rd example
A third unusual phenomena of DST fluids is that a sphere sinking in the fluid  has an oscillating velocity, rather than monotonically approaching a terminal velocity \cite{KSLM11}. It was shown that this also cannot be described with any generalized Newtonian function $\tau(\dot\gamma)$, rather it can also be described in principle by hysteresis in $\tau(\dot\gamma)$  \cite{KSLM11, KSM13}. 

 %relaxation context-model
Making use of this knowledge that hysteresis in $\tau(\dot\gamma)$ is required to explain transient and dynamic phenomena of DST fluids, a simple phenomenological model  was proposed by Ozgen et al. \cite{OBK15}. It consists of a $\tau(\dot\gamma)$ relationship with an effective viscosity that increases with shear rate to mimic shear thickening. This term was made to have hysteresis in $\tau(\dot\gamma)$ such that it depends not only on the instantaneous shear rate, but on a weighted average of shear rate over a preceding time interval, corresponding physically to a time delayed response before the strong solid-like response to impact, as well as a relaxation time over which the effective viscosity decays after the shear rate decreases.  This model was able to qualitatively reproduce the phenomena that were previously argued to require hysteresis; the stable holes in a vibrated layer, and oscillations in the velocity of a sinking sphere. The model also reproduced some phenomena that are known to occur in DST fluids, but have not been explained previously; in particular the abilities of a sphere to bounce and roll on the surface  of the suspension \cite{youtube_bowlingball}. This success is remarkable in that no simulation has been able to produce any one of these phenomenon before -- even individually -- yet several were produced at once with this model. However, this model was made before any of the relevant rheological parameters were measured for real materials, so the parameters were freely tuned to reproduce these phenomena. 

% open problem -  previous work
Despite the popular interest in these transient and dynamic phenomena,  it is remarkable that there has been little systematic study of transient rheology that goes beyond the traditional steady-state $\tau(\dot\gamma)$ function for DST suspensions.    Oscillatory rheology is often used to characterize time-dependence in the constitutive relation.  Typically these results supported the steady shear rheology description \cite{LW03, RK97},  but in general these oscillatory measurement  have  not revealed a different relaxation behavior or any additional features of a constitutive relation that would help explain any transient  phenomena.  There have been some  measurements identifying hysteresis when the control parameter is ramped quickly \cite{De10, BJ12}, but not  detailed enough to expand on the constitutive relation.  Measurements of the delay in stress response after impact identified by Ozgen et al.~\cite{OBK15} have only recently been reported in another paper by us \cite{MMASB17}.  The other  time-dependent behavior that  could result in hysteresis in $\tau(\dot\gamma)$ in combination with the delay before the solid-like like response is the relaxation of the solid-like state \cite{OBK15}.   While a relaxation of stress to a steady-state behavior has been observed  previously \cite{LGHNSPC15},  analysis of this transient, or any trends  in control parameters such as packing fraction,  have not been reported.
 
% detailed problem statement
   In this paper, we characterize the relaxation behavior in rheometer experiments.    To  systematically characterize a relaxation behavior  we report  time series of the  relaxation of stress or shear rate over time after a flow is stopped or stress removed, respectively.   It remains  to be seen  over what range of packing fractions the relaxation behavior can be described by a generalized Newtonian model, and  if other relaxation behaviors are observed which might help explain some of the unusual phenomena  observed in DST suspensions, so we report measurements as a function of packing fractions near the liquid-solid transition where shear thickening behavior is strongest \cite{BJ09, BZFMBDJ11}.    It is  expected that  this relaxation data will be  an essential element in constitutive models of transient and dynamic phenomena such as proposed by Ozgen et al. \cite{OBK15}.

%summary
The remainder of this paper is organized as follows.  We  describe the materials and general methods used in Secs.~\ref{sec:materials} and \ref{sec:methods}, respectively. In Sec.~\ref{sec:viscosity}  we show  typical steady-state   viscosity curves for suspensions of cornstarch and water  from which we obtain  parameter values to compare to transient measurements.  In Sec.~\ref{sec:effective_phi}  we present a method to more precisely characterize the effective weight fraction,  which is helpful for resolving trends in relaxation behavior near the liquid-solid transition.  In Sec.~\ref{sec:relaxation},  we report time series of the  relaxation of stress and shear rate over time.   We identify the different  qualitative types of relaxation observed in stress-controlled experiments  at different  weight fraction ranges  and compare them to a generalized Newtonian model in Secs.~\ref{sec:calibration}-\ref{sec:phi_ranges_stress_control}.  We compare effective viscosities from transient and steady state measurements in Sec.~\ref{sec:viscosity_comparison}.    To  compare to  relaxation in a different type of flow, we  report measurements of shear-rate-controlled relaxation behavior in Secs.~\ref{sec:rate_control}-\ref{sec:calibration_rate_control}.  In Sec.~\ref{sec:timescales}, we  compare the measured relaxation times between stress- and rate-controlled measurements. In Sec.~\ref{sec:discussion}, we compare the results  of different sections, and develop a self-consistent explanation of the relaxation behavior in the range where the relaxation is inconsistent with the generalized  Newtonian model.

\section{Materials}
\label{sec:materials}

Cornstarch was purchased from \textit{Carolina Biological Supply} and suspended in tap water,  to obtain a typical DST fluid \cite{BJ14}. The samples were created at a temperature of $22.0\pm0.6$ $^{\circ}$C and humidity of $48\pm6\%$, where the uncertainties represent day-to-day variations in the respective values.  A four-point scale was used to measure quantities of cornstarch and water to obtain a weight fraction $\phi_{wt}$. While a weight fraction by volume is more traditional, obtaining the weight fraction by volume requires knowing the water content (which depends on temperature and humidity) and porosity of the cornstarch -- both of which are difficult to obtain \cite{BJ14}.  As a result of this difficulty,  volume or weight fraction measurements  made in different laboratories or different seasons with different environmental conditions are generally not comparable, but still useful as a quantitative control parameter within data series from the same lab and season.  Since a volume or weight fraction that could be compared in different laboratories is desirable, we introduce a method in Sec.~\ref{sec:effective_phi} to  identify an effective weight fraction scale that can be compared from lab to lab  and season to season.

Each suspension was mixed until no dry powder was observed. The sample was further shaken in a \textit{Scientific Instruments Vortex Genie 2} for 30 seconds to 1 minute on approximately 60\% of its maximum power output. We directly measured a density of $1200\pm20$ kg/m$^3$ for a suspension at $\phi_{wt}=57\%$ based on the volume and weight in a graduated cylinder.  If we extrapolate based on the fraction of cornstarch and water using the known density of water, this same value  is consistent with the density for suspensions within the uncertainty for weight fractions from 51\% to 63\%, covering our entire measurement range.

\section{Methods}
\label{sec:methods}

Suspensions were measured in an \textit{Anton Paar MCR 302} rheometer in a parallel plate setup. The rheometer measured the torque $M$ on the top plate  and  angular rotation rate $\omega$  of the top plate.   In  different experiments, either torque or rotation rate could be controlled,  while the other was measured as a response. Flows of DST suspensions in such setups are neither uniform nor constant,  exhibiting  variations in the measured parameter in both space and time,   with fluctuations as large as an order of magnitude  larger than the mean over timescales of about 1 second \cite{LDH03}.  Nonetheless, we can  reproducibly characterize each steady state flow by the mean shear stress and shear rate \cite{BJ12}.  The mean shear stress is given by $\tau = 2M/\pi R^3$ where $R$ is the radius of the sample.  While the mean shear rate varies along the radius of the suspension, the mean shear rate at the edge of the plate is used as a reference parameter, which is given by $\dot{\gamma} = R\omega/d$ where $d$ is the gap height between the plates.  The viscosity of the sample is measured as $\eta = \tau/\dot{\gamma}$ in a steady state.   We also measured the force $F$ on the top plate of the rheometer (upward positive) and report the mean normal stress $\tau_N = F/(\pi R^2)$.  To calibrate the normal stress, we subtract the value obtained in steady-state measurements in the limit of zero shear rate.  The gap height was usually set to $d=1.25$ mm unless  otherwise noted, and allowed to vary by $0.05$ mm from experiment to experiment in an attempt to reduce the uncertainty on the sample radius. The sample radius was $R=25.0\pm0.5$ mm, which results in an 8\% error in the calculated shear stress $\tau$ when the plate radius $R$ is used in calculations.   The experiments were performed at a plate temperature of $23.5\pm0.5$ $^{\circ}$C. A solvent trap was used to slow down the moisture exchange between the sample and the atmosphere. The solvent trap effectively placed a water seal around the  sample, with a lipped lid around the sample and the lips touching a small amount of water contained on the top, cupped, surface of the tool.

\section{Steady state viscosity curves}
\label{sec:viscosity}

\begin{figure}
\centering
\includegraphics[width=0.475\textwidth]{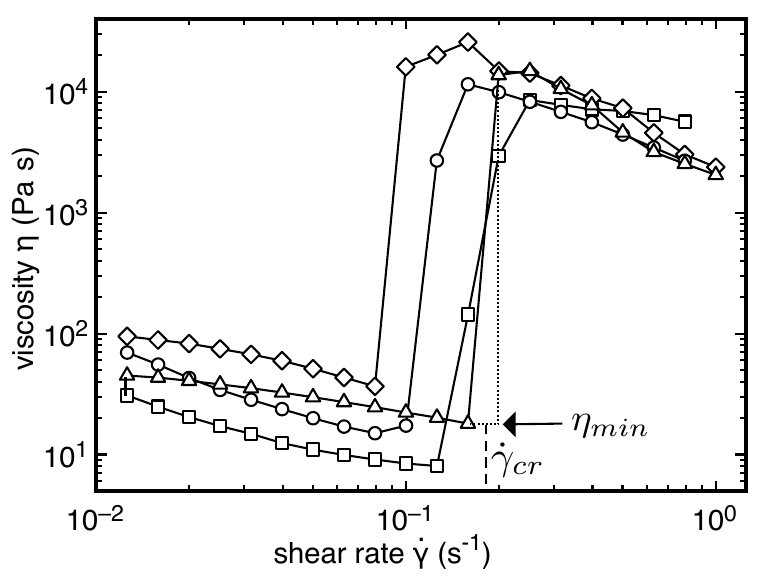}
\caption{An example of repeated ramps of steady-state viscosity $\eta$ vs.~shear rate $\dot\gamma$ for cornstarch and water at weight fraction $\phi_{wt} = 59.6\%$.  The symbols  correspond to different ramps of increasing $\dot\gamma$ (squares, diamonds) and  decreasing $\dot\gamma$ (triangles, circles).  The critical shear rate $\dot{\gamma}_c$ and minimum viscosity $\eta_{min}$ are obtained at the onset of shear thickening for each curve.  %We fit the steepest part of the curve to $\tau \propto \dot{\gamma}^{1/\epsilon}$ (equivalent to $\eta \propto \dot{\gamma}^{1/\epsilon-1}$) to obtain a measure of the difference $\epsilon$ from a discontinuous stress-shear rate relation.
}
\label{fig:viscosity_shearrate}
\end{figure}

While steady-state viscosity curves have been characterized previously  \cite{Ba89, WB09,BJ14}, we present them here for comparison to relaxation measurements in the same samples in later sections. To obtain steady-state viscosity curves, each sample was pre-sheared to produce a state independent of the sample loading history.  We used a linear ramp in shear rate over $200$ seconds, covering a shear rate range that crosses the shear thickening regime and a net shear strain  greater than $100$\%. The shear rate was then ramped down then back up twice with a constant rate of variation on  a logarithmic scale, with $10$ data points per decade and data  averaged for $50$ seconds per point.  

Figure~\ref{fig:viscosity_shearrate} shows an example of the steady state viscosity $\eta$ as a function of shear rate $\dot\gamma$ for the four ramps after the preshear at $\phi_{wt}$=59.6\%.  There is a large run-to-run variation even for the same sample in a constant environment  that seems to be a natural variation; the standard deviation of  the 4 ramps is 30\%, as is typical in measurements of cornstarch and water \cite{BJ12}.  We observed no significant systematic trend  in the repeated ramps,  confirming the pre-shear  eliminated any effects of loading history, environmental change, or any other systematic effects during measurement of a single sample.

\begin{figure}
\centering
\includegraphics[width=0.475\textwidth]{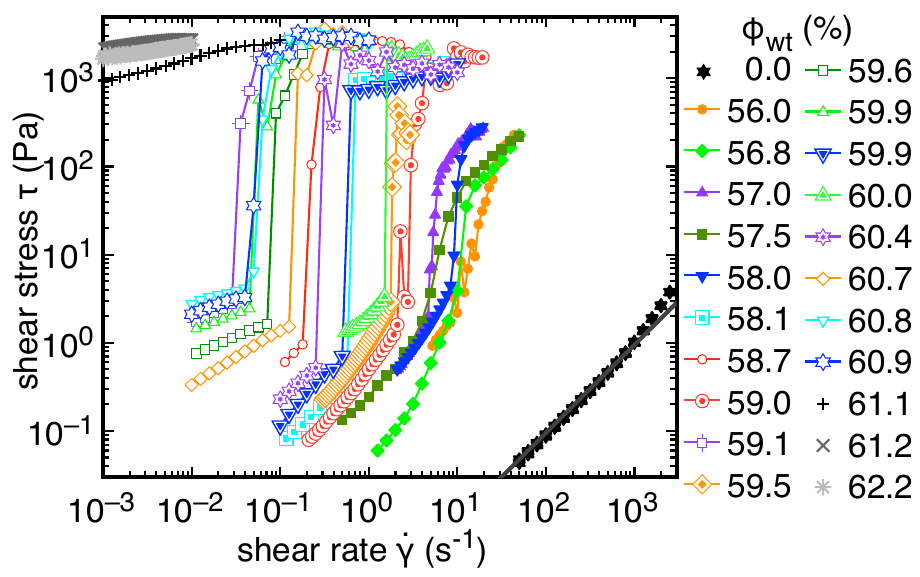}
\caption{(color online) Shear stress $\tau$ as a function of shear rate $\dot{\gamma}$ at  different weight fractions $\phi_{wt}$  shown in the key.    For $58.1\% \le \phi_{wt}<\phi_c=61.0\%$, discontinuous shear thickening is seen as a sharp jump in stress.  For $56.0\% \le \phi_{wt} \le 58.0\%$, we find continuous shear thickening.  %For $\phi_{wt}>\phi_c=61.0\%$, a yield stress of $\sim 10^3$ Pa is observed instead of shear thickening.
}
\label{fig:stress_shearrate_all}
\end{figure}

Averages  of the shear stress $\tau$ as a function of shear rate $\dot\gamma$ over the 4 ramps are shown in Fig.~\ref{fig:stress_shearrate_all}  for different weight fractions $\phi_{wt}$.   At  each weight fraction in the range $58.1\% \le \phi_{wt}< 61.0\%$,  a discontinuous jump in $\tau(\dot\gamma)$ can be seen from $\tau_{min}$ ($\sim 10^0$ Pa) to $\tau_{max}$ ($\sim 10^3$ Pa),  corresponding to discontinuous shear thickening.  In these cases, the jump in shear stress between adjacent points in shear rate was between 1-3 orders of magnitude for each curve.   
%Typically, higher weight fractions have lower critical shear rates $\dot\gamma_c$, but there are some exceptions to this due to the uncertainty in weight fraction when it is measured directly.  This sample-to-sample uncertainty  can come from a combination of variations in the temperature and humidity  within the range stated in Sec.~\ref{sec:materials}, and uncertainties introduced in the process of loading the sample onto the rheometer from an inhomogeneous sample.  
For  $56.0\% \le \phi_{wt} \le 58.0\%$,  some of the repetitions had apparently discontinuous $\tau(\dot\gamma)$ curves, with jumps in shear stress between adjacent points by a factor of 5-10, while others have a more gradual increase,  resulting in average curves that are steep, but  not discontinuous.  Thus, we can define a transition between continuous and discontinuous shear thickening in rate-controlled experiments at $\phi_{wt}$ between 58.0\% and 58.1\%.
%{\bf move?   In this lower weight fraction range, $\tau_{max}$ can be seen in Fig.~\ref{fig:stress_shearrate_all} to be closer to $10^2$ Pa, and the shear stress increase above $\tau_{max}$ is more linear in shear rate, in contrast with the plateau in shear stress seen for higher weight fractions. }

%yield stress
At higher weight fractions of $\phi_{wt} > 61.0$\%, we observed a yield stress on the order of $10^3$ Pa in Fig.~\ref{fig:stress_shearrate_all}. We identified this liquid-solid transition $\phi_c=61.0$\% (also called the jamming transition). We did not observe  shear thickening at weight fractions above $\phi_c$  because of the large yield stress \cite{BJ09, BZFMBDJ11}.

%reynolds number
In all of these cases we present for suspensions, the Reynolds Number $Re = \rho d^2 \dot{\gamma}/\eta < 1$  over the entire shear thickening range. Hence, inertial displacement does not contribute significantly to energy dissipation, and the corresponding dissipative force (proportional to velocity squared) is negligible in relaxation experiments starting from any of these steady states.

\section{Characterizing a more precise effective weight fraction near the critical point}
\label{sec:effective_phi}

%motivation
For a generalized Newtonian fluid, the  energy dissipation rate, and thus the rate of  relaxation is expected to scale with the viscosity  (explained in Sec.~\ref{sec:calibration}), which diverges as it approaches the critical point  at weight fraction $\phi_c$ \cite{BJ09, BZFMBDJ11}.  If  we want to test this scaling close to the critical point,  there is a resolution limit due to the uncertainty in the weight fraction of typically 1-2\%  in most measurements of suspensions.  This uncertainty is due in part to the  variability  of the weight fraction with  temperature and humidity because cornstarch adsorbs water from the atmosphere \cite{BJ12}, and uncertainties introduced in the process of loading the sample onto the rheometer from an inhomogeneous sample. This can lead to large  changes in the measured viscosity in repetitions of experiments from day-to-day and from lab-to-lab,  with infinite sensitivity to the uncertainty in weight fraction due to the divergence of the viscosity at a nearby critical weight fraction $\phi_c$ \cite{BJ09, BZFMBDJ11}. For example, while the critical shear rate $\dot\gamma_c$ at the onset of shear thickening (i.e.~minimum shear rate of the shear thickening range) tends to decrease with weight fraction, in Fig.~\ref{fig:stress_shearrate_all} we  can find counterexamples such as $\phi_{wt}=60.9\%$ and $\phi_{wt}=60.4\%$ with an order of magnitude increase in $\dot{\gamma}_c$ when the apparent weight fraction increases by 0.5\% when $\phi_{wt}$ is within 1\% of $\phi_c$.

%solution
This resolution limit can be circumvented,  and scalings  tested closer to the critical point, by instead characterizing the material in terms of a measurable property that diverges at $\phi_c$, as errors in the diverging quantity will result in much smaller errors in weight fraction.  Two such properties include the minimum viscosity $\eta_{min}$ and the inverse of the critical shear rate $\dot\gamma_c^{-1}$, both measured at the onset of shear thickening \cite{BJ09, FBOB12}.  Thus, in order to  obtain a more reliable measure of effective weight fraction  near the critical point, we measure these values for each $\phi_{wt}$ to use as references for  an effective weight fraction $\phi_{eff}$ that is more  accurate in identifying the sample than $\phi_{wt}$.  

%obtaining Critical shear rate and minimum viscosity
 We identify the critical shear rate $\dot\gamma_c$ and viscosity at the onset of shear thickening $\eta_{min}$ from the viscosity curves in Fig.~\ref{fig:stress_shearrate_all}.   We averaged values of $\dot\gamma$  over the point just before and the point just after the jump shown for example in Fig.~\ref{fig:viscosity_shearrate} for each ramp, then further averaged over the 4 ramps.  The  viscosity at  the lower end of the shear thickening range $\eta_{min}$ was obtained by taking the $\eta$ value just before each jump, then averaging over the four ramps. For both $\dot\gamma_c$ and $\eta_{min}$  the run-to-run variation can be characterized by the standard deviation of the values from the four ramps. In following plots we use the  standard deviation of the mean as the error when fitting mean values in each case, which are on average 16\% for $\dot{\gamma}_{c}$ and 30\% for $\eta_{min}$. We choose to use $\dot\gamma_c^{-1}$ as a reference parameter, having the smaller run-to-run  variation of the two options.

\begin{figure}
\centering
\includegraphics[width=0.475\textwidth]{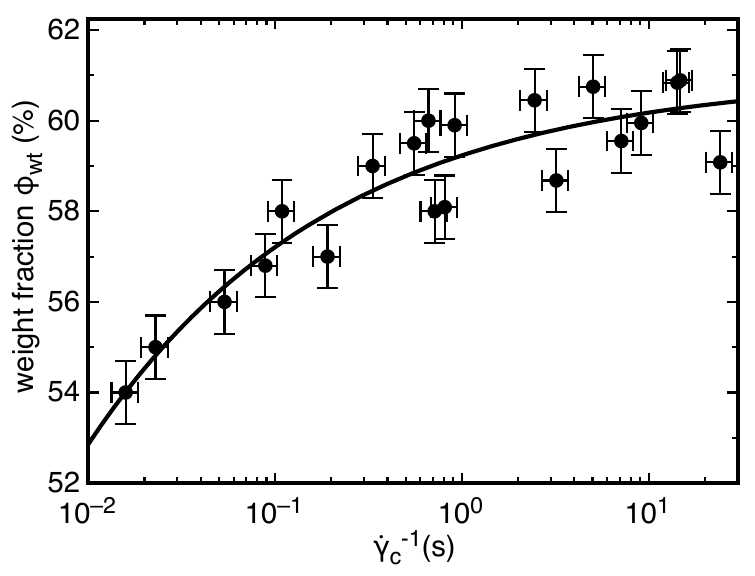}
\caption{Directly measured weight fraction $\phi_{wt}$ as function of  inverse critical shear rate $\dot\gamma_c^{-1}$.   A fit yields a conversion function $\phi_{eff} = -1.8\% \dot{\gamma}_{c}^{0.33} + 61.0\%$,  which can be used to identify an effective weight fraction $\phi_{eff}$  based on the measured  parameter $\dot\gamma_c$,  which is useful for precise characterizations near  the critical weight fraction $\phi_c=61.0\%$.
}
\label{fig:phi_cricshearrate}
\end{figure}

 To obtain a conversion to effective weight fraction $\phi_{eff}$, we plot the  directly measured $\phi_{wt}$  as a function of $\dot\gamma_c^{-1}$ for data  at different weight fractions in Fig.~\ref{fig:phi_cricshearrate}.  The conversion is obtained by least squares fitting a power law $\phi_{wt} = A\dot\gamma_c^{B}+\phi_c$  to the data with fixed $\phi_c=61.0\%$ and fit parameters $A$ and $B$.  We use the 16\% standard deviation of the mean in $\dot{\gamma}_c$ as an input error.  We also adjust errors in $\phi_{wt}$ to a constant value of 0.7\% to obtain a reduced $\chi^2= 1$.  The input error of 0.7\% indicates a combination of the sample-to-sample uncertainty on weight fraction for our measurements plus any deviation of the fit function from the `true' function describing the data. The fit yields $A=-1.8\pm0.2$ and $B=0.33\pm 0.03$,  corresponding to the conversion function
\begin{equation}
\phi_{eff} = -1.8\% \dot{\gamma}_{c}^{0.33} + 61.0\% \ .
%\phi_{eff} = -1.7896\% \dot{\gamma}_{c}^{0.3283} + 61\%
\label{eqn:phi_eff}
\end{equation}
with $\dot\gamma_c$ in units of s$^{-1}$. If  we instead  additionally fit the value of $\phi_c$, then we obtain $\phi_c=60.8\pm 0.6$,  consistent with the value obtained from yield stress measurements for the same data set, and the same uncertainty on weight fraction up to one point in the last digit \cite{BJ09}.

% application of conversion function
With this conversion function, we can now calculate an effective weight fraction $\phi_{eff}$ from a measured $\dot\gamma_c$ with  higher resolution than direct weight fraction measurements. The value of $\phi_{eff}$ using Eq.~\ref{eqn:phi_eff} corresponds to the value of the weight fraction  in our reference temperature  and humidity environment.  

\begin{figure}
\centering
\includegraphics[width=0.475\textwidth]{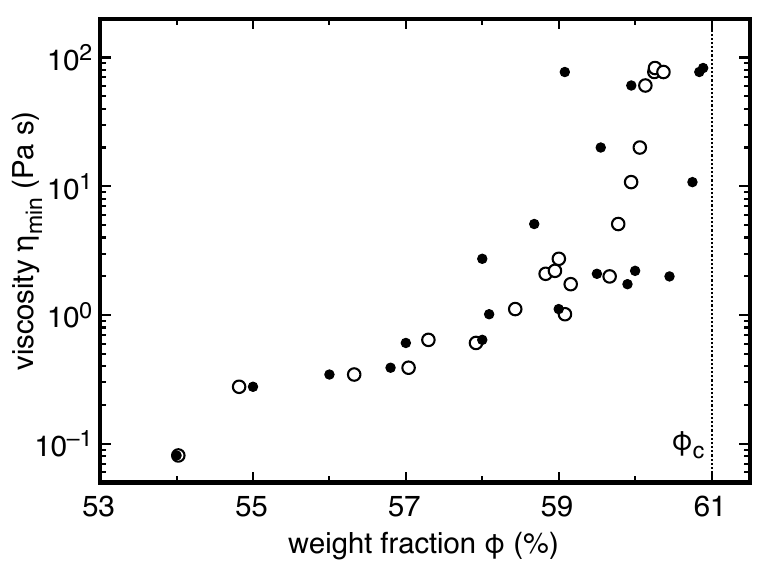}
\caption{Onset viscosity $\eta_{min}$.  Solid symbols:  as a function of the directly measured weight fraction $\phi_{wt}$.  Open symbols:  as a function of the effective weight fraction $\phi_{eff}$ obtained from the fit of  $\dot{\gamma}_{c}^{-1}$ in Fig.~\ref{fig:phi_cricshearrate}.  Using the effective  weight fraction $\phi_{eff}$  results in less scatter, allowing more precise characterization of trends in measured parameters near the critical point.
}
\label{fig:viscosity_phi}
\end{figure} 
 
To test the usefulness of this effective weight fraction $\phi_{eff}$, values of the onset viscosity $\eta_{min}$ from the same measurements as data in Fig.~\ref{fig:phi_cricshearrate} are plotted in Fig.~\ref{fig:viscosity_phi}  for different directly measured weight fractions $\phi_{wt}$.  Near $\phi_c$,  the data is very scattered as expected due to the error on the x-axis,  making it difficult to track the expected divergence in weight fraction.    For comparison,  values of $\eta_{min}$ are plotted in the same figure as a function of $\phi_{eff}$ using Eq.~\ref{eqn:phi_eff} to get $\phi_{eff}$ from  the measured $\dot\gamma_c$ at  each weight fraction.   It can be seen that there is much less scatter in the data in terms of effective weight fraction $\phi_{eff}$ near $\phi_c$.  A power law fit of $\phi_{eff}(\eta_{min})$ analogous to Eq.~\ref{eqn:phi_eff} with the 30\% standard deviation of the mean error on $\eta_{min}$ requires a 0.3\% error on $\phi_{eff}$ to obtain a reduced $\chi^2=1$.  This smaller 0.3\% error on $\phi_{eff}$ than the 0.7\% error on $\phi_{wt}$ confirms that the $\phi_{eff}$ scale based on  the fit of Eq.~\ref{eqn:phi_eff} more precisely relates  to mechanical properties  that diverge near $\phi_c$ (i.e.~$\eta_{min}$ and $\dot\gamma_c^{-1}$) than  direct weight fraction $\phi_{wt}$ measurements.

\section{Relaxation}
\label{sec:relaxation}

\subsection{Relaxation of a generalized Newtonian fluid}
\label{sec:calibration}

%motivation
To measure relaxation times, we performed transient experiments.   We first  applied a constant  torque to the top plate  until the flow reached steady-state,  then removed the applied torque (i.e.~set the controlled shear stress $\tau=0$).  We then  observed the relaxation of the tool over time in terms of the shear rate $\dot\gamma$ at the edge of the plate, as the momentum of the sample and tool was dissipated by the effective viscosity of the fluid.    

%Unlike pure Newtonian fluids, generalized Newtonian fluids do not require a constant viscosity, nor does the relaxation have to be an exponential decay. However, an effective viscosity can still be defined for a generalized Newtonian fluid. We will use the theory to obtain a calculation for a viscosity $\eta_t$ from transient  relaxation measurements to  compare to the viscosity $\eta$ obtained from steady-state viscosity curves. This comparison allows us to study whether the measured viscosity for a sample in steady state is intrinsic to the material, or is there any other quantity that may be intrinsic and is dominant in the relaxation behavior.

% generalized Newtonian fluid theory
To better understand results for suspensions, we compare to a theory for a generalized Newtonian fluid, in which the function $\tau(\dot\gamma)$  obtained from steady-state measurements is expected to be sufficient to describe the relaxation behavior in a transient flow.  While general solutions for the relaxation of the tool for an arbitrary $\tau(\dot\gamma)$ are not necessarily tractable, we can express simple equations and solutions for special cases that approximate relevant cases.  Since steady-state viscosity curves for DST fluids can often be approximated by two Newtonian-like ranges with constant viscosity  separated by the critical shear rate  (Fig.~\ref{fig:stress_shearrate_all}), we start by writing the equations of motion for the special case of a constant viscosity, so that we can later stitch the solutions together.  In a transient flow, the torque $M$  balances the rate of change of angular momentum.  In a stress-controlled flow  where both the tool and the fluid remain rotating during the relaxation, this is equal to the angular acceleration $\dot\omega$ times the sum of the moment of inertia of the fluid $I$ and the moment of inertia of the tool $I_{tool}$.  Since $\dot\omega$ is not uniform in the fluid, we present equations for a characteristic $\dot\omega$ at the edge of the top plate, and the stress assuming a linear flow profile, but this makes the relationships only true as scaling relationships with an unknown proportionality constant.  We will later  make the relationships exact by calibrating with a Newtonian fluid.   In the special case of a constant viscosity over a wide range of shear rates with a laminar flow, the torque simply relates to the viscosity by $M=\eta\omega\pi R^4/2d$ for a circular parallel plate flow geometry \citep{MJ12},   to obtain the differential equation of motion
\begin{equation}
(I + I_{tool})\dot\omega \propto  M = -\frac{\eta\pi R^4\omega}{2d}
\label{eqn:ode}
\end{equation}
This equation has the solution of an exponential decay for $\omega$.  Since the stress $\tau\propto M\propto \omega \propto \dot\gamma$  for a constant viscosity, then the solution for the stress or shear rate is also an exponential decay for a constant viscosity, given by
\begin{equation}
\dot\gamma=  \dot\gamma(t=0) \exp(-t/T_N) \ .
\label{eqn:odesolution}
\end{equation}
 The corresponding timescale of the  exponential decay $T_N$ can be obtained from Eq.~\ref{eqn:ode} to be 
\begin{equation}
T_{N} = \left| \frac{\omega}{\dot\omega} \right| \propto \dfrac{2d}{\pi R^4 \eta}\left( I+I_{tool}\right).
\label{eqn:Trmodel}
\end{equation}
Rearranging Eq.~\ref{eqn:Trmodel} and  substituting $I=\pi\rho R^4 d/2$ for  the moment of inertia of the sample between  circular parallel plates yields  an expression for the viscosity based on the measured relaxation time

\begin{equation}
\eta \propto \dfrac{d}{T_N}\left(\rho d + \dfrac{2 I_{tool}}{\pi R^4}\right).
\label{eqn:transientviscosity}
\end{equation}

% calculation of T_r
\begin{figure}
\centering
\includegraphics[width=0.475\textwidth]{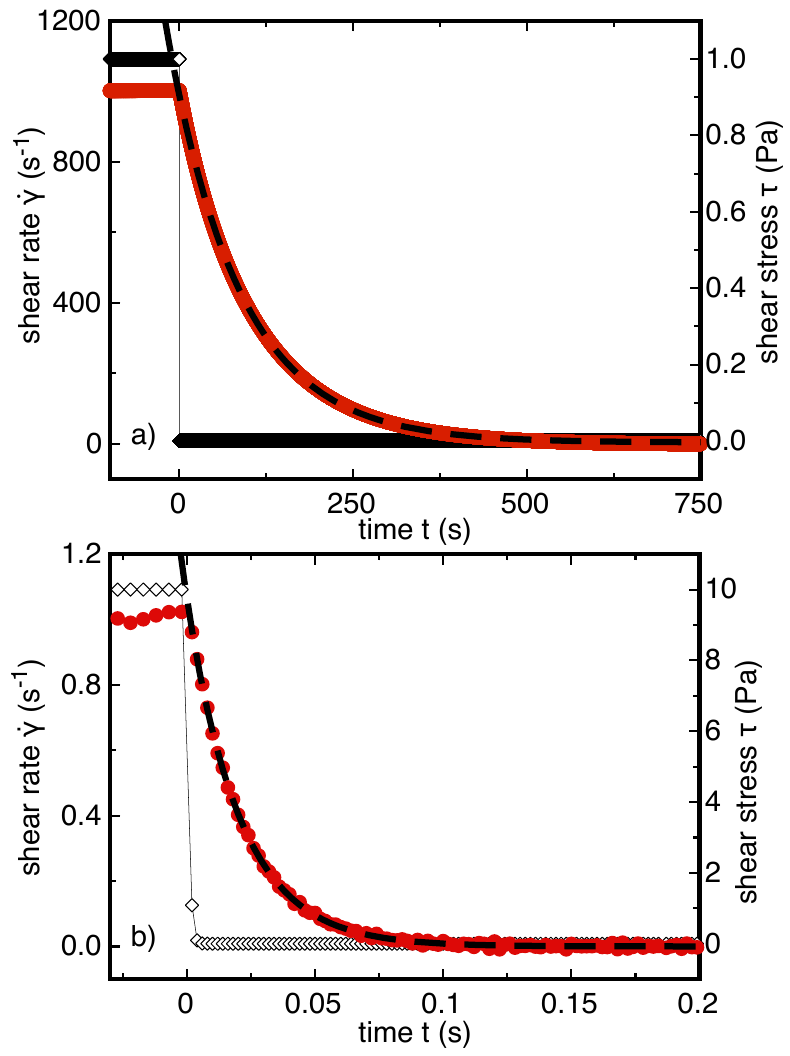}
\caption{(color online) The transient experiment used to  measure a relaxation time $T_N$ for a Newtonian fluid of known viscosity to calibrate Eq.~\ref{eqn:transientviscosity}. (a) water. (b)  silicone oil.  Black open diamonds:  the controlled stress level $\tau$,  which was set to zero at $t = 0$ s after a steady state flow (right axis scale).  Red filled circles:  measured shear rate $\dot\gamma$ (left axis scale). Dashed line:  exponential fit to obtain the relaxation timescale $T_N$.
}
\label{fig:watercalibration}
\end{figure}

%calibration methods
To calibrate the proportionality coefficient in Eq.~\ref{eqn:transientviscosity}, we used water as a Newtonian fluid with known viscosity $\eta$, and measured the relaxation time $T_N$ in a stress-controlled relaxation experiment.  To ensure a  laminar flow, we set a smaller gap of $d = 0.550$ mm for this experiment.   We also measured a steady state viscosity curve (shown in Fig.~\ref{fig:stress_shearrate_all} as $\phi_{wt}=0\%$)  to find the maximum shear stress where it remains proportional to shear rate  before  the flow becomes non-laminar due to turbulence at high Reynolds numbers. We show a linear fit to the water data for low shear rate where the flow remains laminar.  Note that while the slope of the viscosity curve increases with shear rate at stresses above 1 Pa,  this consequence of turbulence is not usually referred to as shear thickening.  We then use this value of 1 Pa as the initial stress set point value in the  stress-controlled relaxation experiments for water to maximize the stress resolution during relaxation of a laminar flow.   

%relaxation experiment
For the  calibration experiment to measure the relaxation behavior of water, we set $\tau=1$ Pa until the flow reached a steady shear rate $\dot\gamma$, then  set $\tau=0$  at a time defined to be $t=0$.   Figure \ref{fig:watercalibration}a shows  an example of the relaxation for  one of these experiments.  After the steady behavior for $t<0$, there is a  gradual decay in the  measured shear rate to zero. To obtain $T_N$, we fit the general solution to the differential equation (Eq.~\ref{eqn:odesolution}) to the data once the measured stress reached the set point value for $t>0$.  The input error on the fit was adjusted to obtain a reduced $\chi^2\approx1$, requiring an input error of only 0.5\% of the peak shear rate.  This small difference of 0.5\% between the fit and the data  confirms the model of Eqs.~\ref{eqn:ode} and \ref{eqn:odesolution} describes the  relaxation of Newtonian fluids well.

%calibration results
The calibration  now allows us to determine the proportionality coefficient in Eq.~\ref{eqn:transientviscosity}.  Using the measured  relaxation time of $T_N=105.05 \pm 0.03$ s, the measured  steady-state viscosity $\eta=9.1 \times 10^{-4}$ Pa s of  the same sample of water  (3\% smaller than the nominal value of $9.4 \times 10^{-4}$ Pa s \cite{Wh09}), and the measured moment of inertia  of the tool $I_{tool} = 1.282\times10^{-5}$ kg m$^{2}$,  the  proportionality coefficient in Eq.~\ref{eqn:transientviscosity} is calculated to be 8.1.  This  calibration makes the proportionality of Eq.~\ref{eqn:transientviscosity} exact,  which can be used to obtain a  measure of viscosity $\eta_t$ from transient stress-controlled relaxation flows from the relaxation time $T_1$ we measure for suspensions

\begin{equation}
\eta_t = 8.1 \dfrac{d}{T_1}\left(\rho d + \dfrac{2 I_{tool}}{\pi R^4}\right) \ .
\label{eqn:transientviscosity_calib}
\end{equation}

%silicone oil
To confirm the calibration on a shorter timescale, we performed a second transient relaxation experiment for a Newtonian fluid using silicone oil (nominal viscosity of 10,000 cSt at 25$^\circ$C, density $\rho = 971$ kg/m$^3$).  We report results in Fig.~\ref{fig:watercalibration}b for the transient experiment with silicone oil with a gap $d = 1.220$ mm. We fit Eq.~\ref{eqn:transientviscosity} to the data to obtain a relaxation time of $T_N = 0.0207 \pm 0.0001$ s, where the input error in the fit was adjusted to 0.5\% of the peak shear rate to obtain a reduced $\chi^2 \approx 1$, just as good a fit as for the water data.  We measured the steady state viscosity to be 9.9 Pa s.  These values give a calibration coefficient  6\% smaller than for water in Eq.~\ref{eqn:transientviscosity}. This difference is smaller than the 8\% sample-to-sample variation in viscosity measurements due to the uncertainty in $R$, so remains small compared to other errors and comparisons reported in this paper.  This confirms that the equipment used can reliably resolve the transient behavior for relaxation times as short as 0.21 s, shorter than all but one of the measured relaxation times reported later in this paper.

\subsection{Relaxation in stress-controlled experiments}
\label{sec:stress_control}
%motivation and methods for stress control
To test the applicability of the generalized Newtonian model and measure a relaxation time $T_1$ for suspensions, we initially set the shear stress $\tau$ controlled by the rheometer to a constant stress just above  $\tau_{max}$ (within 10\%), where $\tau_{max}$ is the maximum stress in the shear thickening range from steady state viscosity curves as shown in Fig.~\ref{fig:stress_shearrate_all}. After steady state was reached, we set the control to $\tau=0$ at a time which was then defined to be $t=0$. We then observed the relaxation of $\dot\gamma$, corresponding to the shear rate at the outer edge of the top plate, as the inertia of both the tool and the fluid was dissipated.

\begin{figure}
\centering
\includegraphics[width=0.475\textwidth]{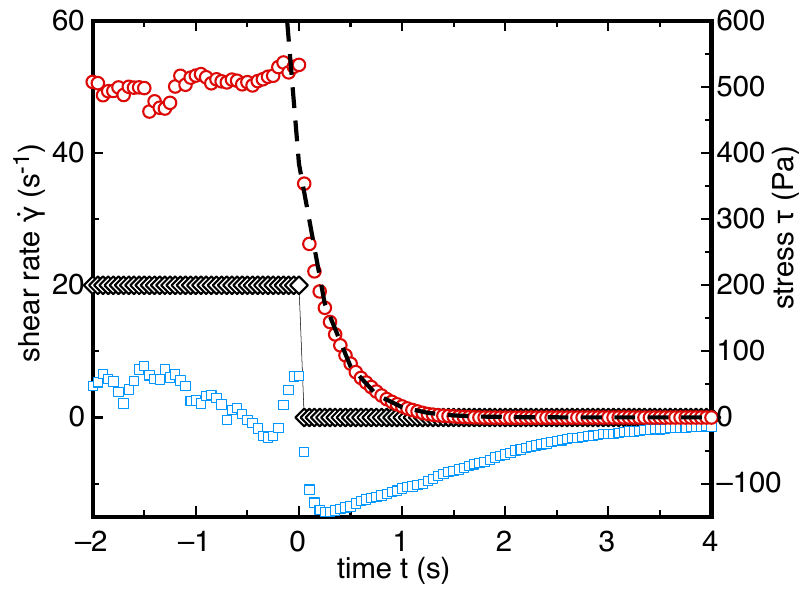}
\caption{(color online) An example of a transient stress-controlled experiment used to  measure a relaxation time  in the generalized Newtonian regime at low weight fraction ($\phi_{eff} = 57.1$\%). Black diamonds:  the controlled shear stress $\tau$ (right axis scale). Red circles:  measured response in shear rate $\dot\gamma$ (left axis scale). Blue squares: measured normal stress (right axis scale).  Dashed line:  exponential fit to $\dot\gamma$ obtain the relaxation timescale $T_1$. This shear rate relaxation is consistent with the generalized Newtonian model,  with an initially fast relaxation due to the larger steady-state viscosity  for $\dot\gamma\stackrel{>}{_\sim}\dot\gamma_c$,  followed by an exponential decay for $\dot\gamma\stackrel{<}{_\sim}\dot\gamma_c$ where the steady-state  viscosity is nearly constant.
}
\label{fig:shearrate_time_newtonian}
\end{figure}

%results - low phi
In Fig.~\ref{fig:shearrate_time_newtonian}, we show an example of a stress-controlled relaxation experiment at $\phi_{eff} = 57.1$\%, representative of the lower weight fractions  we measured.  While for suspensions there are fluctuations in the steady state for $t<0$, for $t>0$ the shear rate is again observed to decay to zero.   We define the relaxation time $T_1$ for  stress-controlled measurements by fitting this data to an exponential decay given by
\begin{equation}
\dot\gamma \propto \exp(-t/T_1) \ .
\label{eqn:Tr1_stresscontrol}
\end{equation}
Since it took a short period after $t=0$ for the tool to reach its set point stress, we fit data once the recorded shear stress was less than 0.001 Pa, which  generally happened in less than 0.02 s.  The  input error for the fits was adjusted until the reduced $\chi^2 \approx 1$ to obtain an error on the fit value of $T_1$.  For the data in Fig.~\ref{fig:shearrate_time_newtonian}, we find that the relaxation of shear rate is consistent with an exponential within a root-mean-square difference of 0.5\% of the initial stress, indicating an excellent fit of Eq.~\ref{eqn:Tr1_stresscontrol} in this range.   The transient viscosity $\eta_t$ obtained from Eq.~\ref{eqn:transientviscosity_calib} for this value of $T_1$ is within 43\% of the value of  the steady-state viscosity $\eta_{min}$  at the onset of shear thickening for the same sample, comparable to the typical sample-to-sample standard deviation of 40\%, consistent with a Newtonian fluid. However, the proportionality in the fit of Eq.~\ref{eqn:Tr1_stresscontrol} is  significantly lower than the shear rate for $t<0$,  a difference from the Newtonian behavior in Fig.~\ref{fig:watercalibration}.  Since this sample is shear thickening with 2 nearly constant viscosity regions as seen in Fig.~\ref{fig:stress_shearrate_all}, the deviation could be  explained by the steady state viscosity $\eta(\dot\gamma)$ being larger than $\eta_{min}$ for $t < 0$, where $\tau>\tau_{max}$, leading to a much larger relaxation rate  until the  stress dropped below $\tau_{min}$ (or equivalently,  the shear rate dropped below $\dot\gamma_c$).  At  later times,  when $\dot\gamma < \dot\gamma_c$, the steady state $\tau(\dot\gamma)$  curve seen in Fig.~\ref{fig:stress_shearrate_all} is nearly linear,  corresponding to a constant viscosity and an expected  exponential relaxation.  In Fig.~\ref{fig:shearrate_time_newtonian}, the proportionality fit is equal to $2.2\dot\gamma_c$, which is  reasonably near $\dot\gamma_c$. These observations are consistent with a generalized Newtonian model where  the steady-state viscosity curve $\tau(\dot\gamma)$  can describe the relaxation in shear rate for this weight fraction.

%normal stress
 We also show the normal stress measured during the  relaxation experiment in Fig.~\ref{fig:shearrate_time_newtonian}  as blue squares using the right axis scale.    The normal stress is  negative for $t>0$, and relaxes much like the shear rate.  Notably, the normal stress does not track the shear stress closely for $t>0$, in contrast to what is typically found in DST fluids in steady-state flows \cite{LDHH05, BJ12, BJ14}.  While  the generalized Newtonian model in terms of $\tau(\dot\gamma)$ does not make explicit predictions for the normal stress, it  will be insightful to compare with the normal stress at other weight fractions later.

 \begin{figure}
\centering
\includegraphics[width=0.475\textwidth]{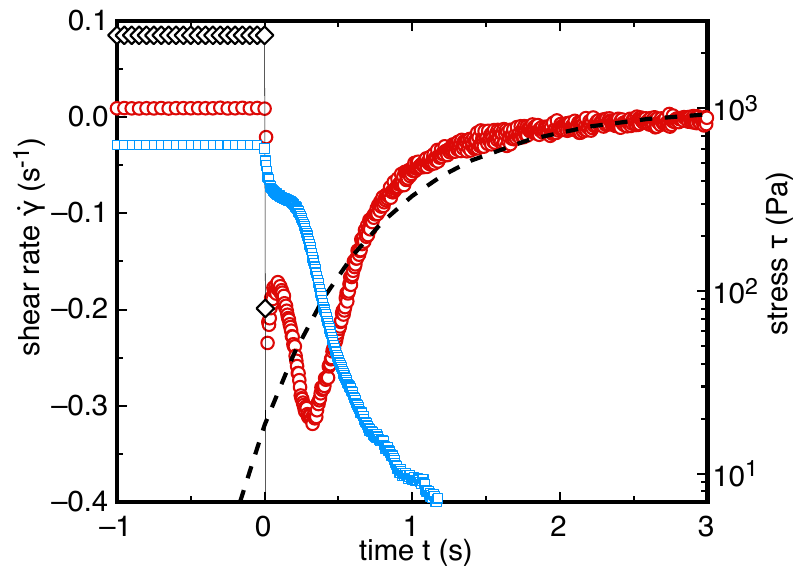}
\caption{(color online) An example of a transient stress-controlled experiment used to measure a relaxation time in the high weight fraction range ($\phi_{eff} = 60.3$\%) below $\phi_c$. Black diamonds:  the controlled shear stress $\tau$ (right axis scale).  Red circles:  measured response in shear rate $\dot\gamma$ (left axis scale). Blue squares: measured normal stress (right axis scale). Dashed line:  exponential fit to $\dot\gamma$ to obtain the relaxation timescale $T_1$. The negative shear rate and its oscillation are inconsistent with the generalized Newtonian behavior seen in Fig.~\ref{fig:shearrate_time_newtonian}.  Along with the positive normal stress, these features are suggestive of a temporary solid-like structure of particles in contact that can store energy elastically.
}
\label{fig:shearrate_time_nonnewtonian}
\end{figure}

% results -high phi
At higher weight fraction, but still below $\phi_c$, we found a different relaxation behavior. Figure \ref{fig:shearrate_time_nonnewtonian} shows an example at $\phi_{eff} = 60.3$\%.    In contrast to the data at lower $\phi_{eff}$ in Fig.~\ref{fig:shearrate_time_newtonian}, the shear rate became negative  immediately after $\tau$ was set to zero, at a rate  much larger than the initial shear rate, corresponding to the tool springing backwards.  The springing backwards is  suggestive of some elasticity and energy storage in the sample, such that the initial shear stored energy in strain, which could be released when the applied applied torque was removed from the tool, pushing the tool back.  The  single oscillation observed in  Fig.~\ref{fig:shearrate_time_nonnewtonian} is another characteristic of elasticity, corresponding to an underdamped regime of an oscillator.  This behavior could in principle be described by a generalized Newtonian model with linear viscoelastic term, (similar to Eq.~\ref{eqn:ode}, but with an additional torque $-\pi R^3G\gamma/2$) due to the shear modulus $G$. We estimate  $G \sim 2d(2\pi/T)^2 (I+I_{tool})/\pi R^4 = 7$ Pa, where $T=0.4$ s is the measured period of oscillation in Fig.~\ref{fig:shearrate_time_nonnewtonian}.   This  shear modulus is small compared to $\tau_{max}$ and thus too small to  be noticeable in most measurements of DST fluids.  In the linear viscoelastic model, the viscous term $\eta$ still relates to the timescale of an exponential decay of the oscillation with the same relationship as Eq.~\ref{eqn:transientviscosity}, so to obtain a relaxation timescale $T_1$, we fit the data for $t>0$ to an exponential decay as in Eq.~\ref{eqn:Tr1_stresscontrol}, despite the poor fit.  Despite the ability to fit to a viscoelastic model, the negative shear rate and its oscillation differ qualitatively from the behavior at lower weight fraction shown in Fig.~\ref{fig:shearrate_time_newtonian},  and are likewise not predicted by the generalized Newtonian model based on the measured $\tau(\dot\gamma)$ in Fig.~\ref{fig:stress_shearrate_all}.  
%but has  been observed before  in a superposition of  pure and oscillatory shear \cite{???}.   
% Estimated T from the figures = 0.4 sec, which gives frequency = 2.5 1/s, angular velocity = 15.71  rad/s. I = 9.204e-7 kg m^2, I_tool = 1.282e-5 kg m^2, R = 0.025 m, $d=0.00125$ m

% Normal stress
The  measured normal stress is also shown in Fig.~\ref{fig:shearrate_time_nonnewtonian}.  We only show the normal stress down to the measurement resolution, and it is smoothed over a range of 0.05 s.  In  this case the normal stress was positive (pushing upward on the top plate).  This could be the result of a temporary system-spanning  structure of solid particles in contact,  which could support a load and allow elastic energy storage.   Such  positive normal stresses resulting from system-spanning  contact networks have also been shown to be prominent in steady-state Discontinuous Shear Thickening at high weight fractions \cite{BJ12, SMMD13}.

\subsection{Weight fraction regimes of different stress-controlled relaxation behaviors}
\label{sec:phi_ranges_stress_control}

\begin{figure}
\centering
\includegraphics[width=0.475\textwidth]{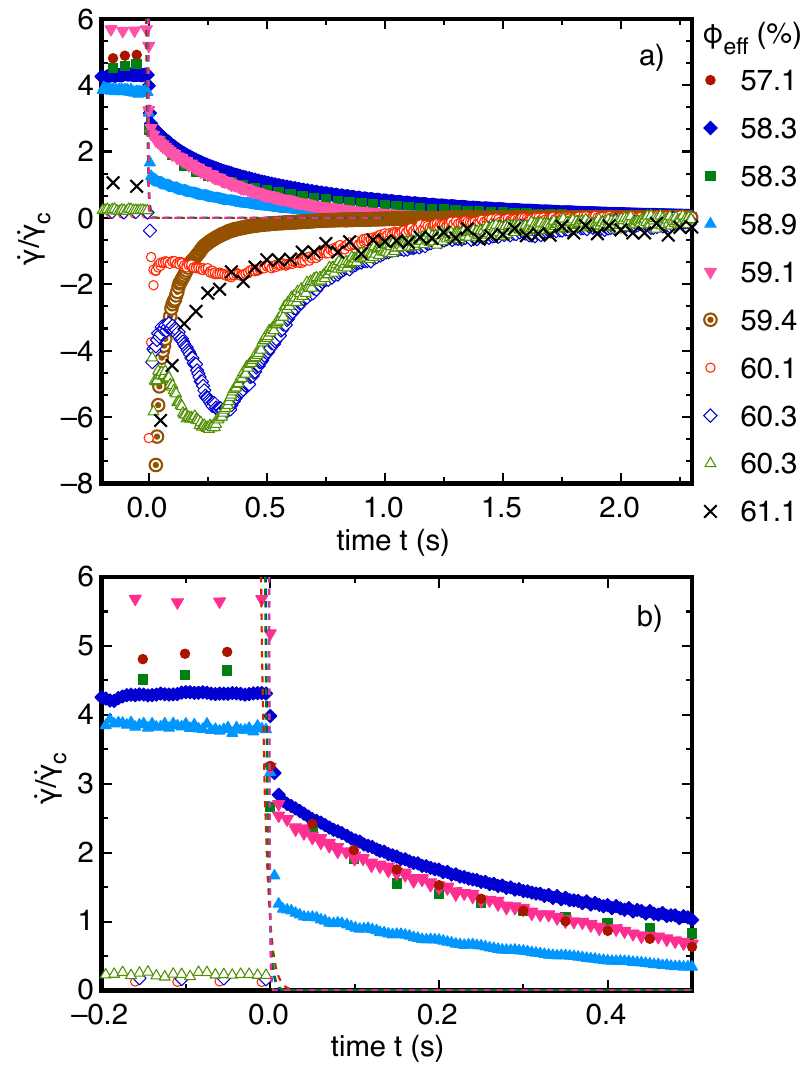}
\caption{(color online) Relaxation of shear rate $\dot\gamma$ over time $t$, normalized by the critical shear rate $\dot\gamma_c$ for different weight fractions $\phi_{eff}$ (listed in the legend).  (a) For $\phi_{eff} > 59.4\%$, we observe negative shear rates after the stress is set to zero, and oscillation for $60.1 \le \phi_{eff} \le 60.3 \%$ (open symbols), qualitatively inconsistent with the generalized Newtonian model. (b) Zoomed scale of panel a to see the faster initial relaxation behavior for $\phi_{eff}\le 59.1\%$. Dashed lines: upper limit of the fast, early relaxation behavior based on the generalized Newtonian model of Eq.~\ref{eqn:transientviscosity_calib} using the maximum viscosity of the shear thickening range $\eta_{max}$,  confirming the fast relaxation is consistent with the generalized Newtonian model in this range. 
} 
\label{fig:shearrate_time_all}
\end{figure}

To identify the range of  weight fractions where the behaviors seen in Figs.~\ref{fig:shearrate_time_newtonian} and \ref{fig:shearrate_time_nonnewtonian} occur, we plot an example of the shear rate relaxation for each weight fraction we measured in Fig.~\ref{fig:shearrate_time_all}a.    The shear rate $\dot\gamma$ is normalized by the critical shear rate $\dot\gamma_c$ for each weight fraction since the initial shear rates were on very different scales due to  the variation of $\dot\gamma_c$ with weight fraction.   Due to the uncertainty of 0.3\% on $\phi_{eff}$, it is still difficult to make out systematic trends of relaxation times within the small regimes with a single qualitative behavior in these plots, but these  plots are still helpful to identify the range of $\phi_{eff}$ where the different qualitative behaviors were found.

%low phi
For samples with $\phi_{eff} \le 59.1\%$, the  relaxation behavior is qualitatively  similar to the generalized Newtonian behavior shown in Fig.~\ref{fig:shearrate_time_newtonian}.  This  can be seen more clearly in Fig.~\ref{fig:shearrate_time_all}b, where we show a zoomed in version of Fig.~\ref{fig:shearrate_time_all}a.  In this range of $\phi_{eff}$, after the tool relaxation and before the exponential decay, the first value of shear rate measured after the tool relaxation (i.e.~the applied shear stress is less than 0.001 Pa) is on average $(2.2\pm 0.5)\dot\gamma_c$ in this range,  where the uncertainty corresponds to the standard deviation of multiple experiments.   We test if this faster relaxation is  consistent with the generalized Newtonian model by plotting  an upper limit on the shear rate $\dot\gamma$ as the dashed lines in Fig.~\ref{fig:shearrate_time_all}b, based on the generalized Newtonian model of Eq.~\ref{eqn:transientviscosity_calib} using the maximum viscosity of the shear thickening range $\eta_{max}$. This is an upper limit on the initial shear rate in the generalized Newtonian model, as the actual viscosity in this range would vary between $\eta_{max}$ and $\eta_{min}$ as the shear rate decreases through the shear thickening range. The fact that these curves  correspond to a faster relaxation than we observe (i.e.~they are to the left of the data)  as far as we can  resolve is at least consistent with the generalized Newtonian model.  

%high phi
At higher weight fractions  ($59.4\% \le \phi_{eff} \le 60.3\%$) in Fig.~\ref{fig:shearrate_time_all}a, the shear rates are negative for $t>0$ like in Fig.~\ref{fig:shearrate_time_nonnewtonian},  inconsistent with the generalized Newtonian model of Eq.~\ref{eqn:ode}.  The  single oscillation is also seen for all of the datasets in the range $60.1\% \le \phi_{eff} \le 60.3\%$.  At $\phi_{eff}=59.4\%$,  the oscillation was not found,  which may be because the relaxation time  ($0.062$ s) is much shorter  than the oscillation period at higher effective weight fractions ($0.4$ s), which  would typically result in the response being  in the overdamped regime of a harmonic oscillator (i.e.~as in a viscoelastic system).

%variability in iniitial shear rate
 To  test which  aspects of the relaxation are independent of the applied stress, we performed a series of experiments with different applied stress for $t<0$ at a fixed $\phi_{eff}$ in the high-weight-fraction range.  We found that  the  qualitative shape of the shear rate relaxation curves varies with the applied shear rate, so the curves shown in Fig.~\ref{fig:shearrate_time_all} should not be  interpreted as the only possible qualitative results.  On the other hand, some common features exist from  which we can draw general conclusions.  In particular, for applied stress $\tau \ge 70$ Pa up to $\tau_{max}$, the relaxation behavior was always qualitatively inconsistent with the generalized Newtonian model, and the normal stress was always negative.  For $\tau_{min} < \tau < \tau_{max}$, the measured relaxation time had no clear trend in $\tau$, but the value varied significantly at different stresses, with a standard deviation of 0.3 decades (i.e.~a factor of 2).

%above phi_c
We also performed measurements at $\phi_{wt}=61.1\%$, shown in Fig.~\ref{fig:shearrate_time_all}a.  Since this is above the liquid-solid  transition at $\phi_c$,  there is no shear thickening and thus no $\tau_{max}$, $\dot\gamma_c$, or values of $\phi_{eff}$.   Thus, we set the initial shear stress based on the value of $\tau_{max}$  from weight fractions just below $\phi_c$, used the steady-state shear rate to normalize $\dot\gamma$,  and give the raw value of $\phi_{wt}$ in the legend in Fig.~\ref{fig:shearrate_time_all}a.  We observe a negative stress for $t<0$ like at the other high weight fractions, but without oscillation. Since this is above the liquid-solid  transition at $\phi_c$,  the  steady-state rheology is dominated by a yield stress.  It is likely that the resulting enhanced dissipation would be enough to prevent energy storage from resulting in oscillation, analogous to overdamping a harmonic oscillator.

\subsection{Comparison between steady state and transient viscosity as a function of weight fraction}
\label{sec:viscosity_comparison}

\begin{figure}
\centering
\includegraphics[width=0.475\textwidth]{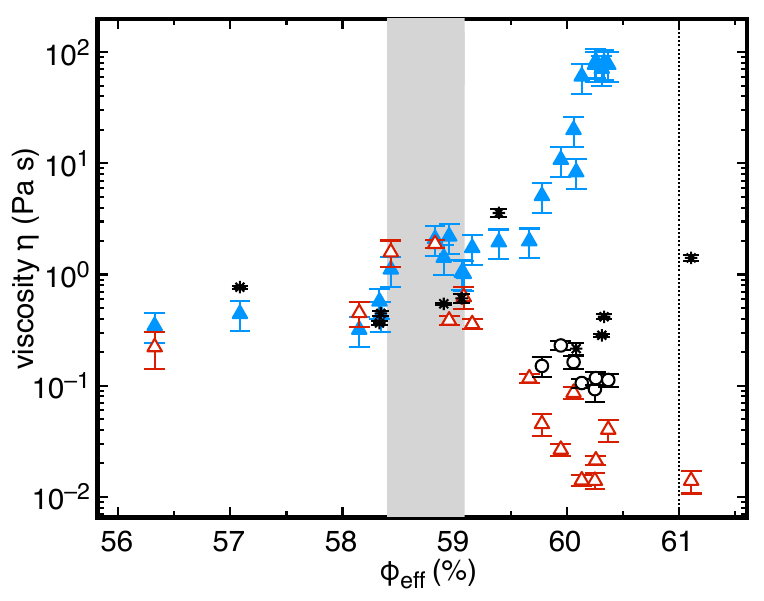}
\caption{(color online) Steady state and transient viscosities as a function of effective weight fraction $\phi_{eff}$. Blue solid triangles: steady state viscosity $\eta_{min}$. Black stars: transient viscosity $\eta_t$ for shear-stress-controlled relaxation experiments. Red open triangles: transient viscosity $\eta_t$ for shear-rate-controlled experiments based on relaxation time $T_1$. Black open circles: transient viscosity $\eta_t$ for shear-rate-controlled experiments based on the relaxation time $T_2$. Vertical dotted line: liquid-solid transition at $\phi_{eff} = \phi_c$. The transient viscosity in both control schemes deviates from the steady state viscosity at large $\phi_{eff}$, by up to 4 orders of magnitude. Vertical gray band:   uncertainty in the transition delineating whether relaxation behavior follows the generalized Newtonian model.
}
\label{fig:viscosity_phi_phic}
\end{figure}

%motiivation
We next quantitatively test how well the generalized Newtonian model describes the relaxation time $T_1$, to determine whether the failure of Eq.~\ref{eqn:Tr1_stresscontrol} to describe the shape of the transient relaxation curves is merely due to a missing non-viscous term in the equation of motion such as an elastic term, or also involves a failure to correctly describe the energy dissipation via the relationship between $T_1$ and $\eta_t$ in Eq.~\ref{eqn:transientviscosity}.  We do this by comparing the transient viscosity $\eta_t$ with the steady state viscosity $\eta_{min}$ as a function of effective weight fraction $\phi_{eff}$ in  Fig.~\ref{fig:viscosity_phi_phic}.

%steady state viscosity
In Fig.~\ref{fig:viscosity_phi_phic}, the solid triangles show $\eta_{min}$ as a best estimate of the hydrodynamic viscosity obtained as in Fig.~\ref{fig:viscosity_shearrate}. The error bars represent the standard deviation of the mean of the four ramps. The steady state viscosity increases with $\phi_{eff}$, with a typical apparent divergence as $\phi_{eff} \rightarrow \phi_c$ \cite{BJ09, BZFMBDJ11}.  

%transietn viscosity - 
The stars in Fig.~\ref{fig:viscosity_phi_phic} represent the transient viscosity $\eta_t$ from the stress-controlled relaxation measurements, calculated from Eq.~\ref{eqn:transientviscosity_calib} where $T_1$  is obtained from the fit of  Eq.~\ref{eqn:Tr1_stresscontrol}. The errors shown are  propagated from the standard deviation of the mean of $T_1$ for 5 repeated measurements of each sample, added in quadrature with the fit errors.  

%comparison- low phi
For our lowest weight fractions $\phi_{eff} \le 59.1$\%,  where the relaxation behavior seen in Fig.~\ref{fig:shearrate_time_all} was qualitatively consistent with the generalized Newtonian model, we find that the transient viscosity $\eta_t$ is scattered around the steady state viscosity $\eta_{min}$. The root-mean-square difference between $\eta_t$ and $\eta_{min}$ is 47\%.  This is larger than the average errors on $\eta_t$ or $\eta_{min}$ for individual samples of 5\% for stress-controlled data and 15\% for rate-controlled data. However, the 47\% difference is  comparable to the sample-to-sample variation in $\eta_{min}$ of 40\%, measured as the root-mean-square difference between the measured $\eta_{min}$ and a best power law fit of $\eta_{min}$ in the same range of $\phi_{eff}$. This agreement within the scatter confirms that the relaxation behavior is consistent with that of a generalized Newtonian fluid in this range of $\phi_{eff}$, where the exponential relaxation rate relates to $\eta_{min}$.
%($\phi_{eff}/\phi_c \le 0.968$)

% one intermediate case - 59.4
At the next lowest weight fraction, $\phi_{eff} = 59.4\%$,  corresponding to the high-weight-fraction range of Fig.~\ref{fig:shearrate_time_all},  $\eta_t$ is  83\%  larger than $\eta_{min}$,  within 2 standard deviations of the scatter  of $\eta_{min}$.  However, the  relaxation behavior was qualitatively inconsistent with that of a generalized Newtonian fluid, since the shear stress dropped negative for $t>0$, with the opposite sign of a Newtonian fluid. Thus,  this weight fraction remains the lowest  at which we find the relaxation behavior  to be inconsistent with the prediction for a generalized Newtonian fluid based on the viscosity curves in Fig.~\ref{fig:stress_shearrate_all}
%($\phi_{eff}/\phi_c = 0.974$)

%transietn viscosity - intermediate phi
At higher weight fractions, $60.1\% \le \phi_{eff} < 61.0\%$, where the qualitative behavior shown in Fig.~\ref{fig:shearrate_time_all} is also  inconsistent with the generalized Newtonian model, we find that the transient viscosity $\eta_t$ is smaller than the steady state viscosity  $\eta_{min}$ by 1.5 to 2.5 orders of magnitude in Fig.~\ref{fig:viscosity_phi_phic}.  We note that accounting for the non-Newtonian  viscosity function $\tau(\dot\gamma)$ cannot reduce this discrepancy; since $\eta_{min}$  is the minimum viscosity in the viscosity curve, the values shown for the steady-state viscosity are the lowest we could have chosen at any given weight fraction, and so the discrepancy would be even larger  if we compared to viscosity values of different shear rates or tried to account for the non-Newtonian shape of $\tau(\dot\gamma)$.  

%phi>phi_c
For $\phi_{wt} > \phi_c$, we could still  measure a relaxation time despite the fact that the material was a solid in the sense that it had a yield stress.  We found a  transient viscosity in between the steady state viscosity and transient viscosity at $\phi_{eff}$ just below $\phi_c$.  Based on the steady-state $\tau(\dot\gamma)$ curve, no flow is expected for $\tau<\tau_{min}$.  If we include strain-dependent rheology as in the viscoelastic model, yield stress materials generally have a stress-strain curve that goes to zero strain in the limit of zero stress. This would lead to an expectation of the tool springing back with a negative shear rate to relax the strained state, as observed in Fig.~\ref{fig:shearrate_time_all}.  However,  the steady-state curves for yield stress fluids have an infinite viscosity in the limit of zero shear rate, and   even at the initial shear rate used, an effective viscosity close to $\eta_{max}$ of the highest $\phi_{eff}<\phi_c$.  That puts a lower bound on the  effective viscosity  for the generalized Newtonian model  on the order of $10^5$ Pa s.  The fact that $\eta_t$ is  on the order of 1 Pa s for $\phi_{wt}>\phi_c$, and much lower than  the values of $\eta_{min}$ for $\phi_{eff}<\phi_c$, is  inconsistent with  the generalized Newtonian model.

\subsection{Relaxation in rate-controlled experiments}
\label{sec:rate_control}

 To test the generality of the observed relaxation behaviors for different types of flows with different boundary conditions, we also performed relaxation measurements in rate-controlled flows.  These experiments were analogous to the stress-controlled experiments with the roles of shear stress and shear rate swapped.  In these measurements, we first rotated the tool at a shear rate $\dot\gamma$ about 60\% higher than the critical shear rate $\dot\gamma_c$  to achieve a steady-state in the stress at a value above the maximum stress $\tau_{max}$ of the shear thickening range (i.e.~in  the high stress regime seen in Fig.~\ref{fig:viscosity_shearrate}). After a steady-state was reached, we attempted to stop the tool  by setting the shear rate to $\dot\gamma=0$,  and measured the relaxation of the stresses on the tool over time due to the relaxation of the fluid.  
 
 \begin{figure}
\centering
\includegraphics[width=0.475\textwidth]{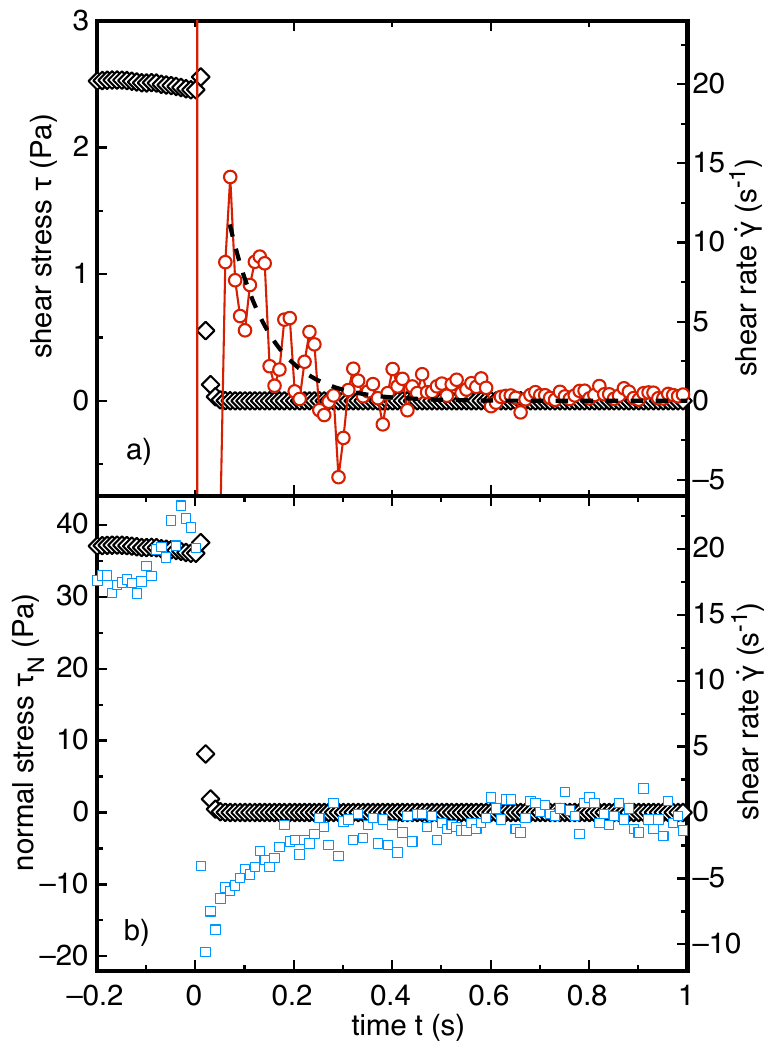}
\caption{(color online) (a) An example of a transient rate-controlled experiment in the generalized Newtonian regime at low weight fraction ($\phi_{eff}=58.1$\%). Black diamonds:  the controlled shear rate $\dot\gamma$ (right axis scale), which was set to zero for $t>0$ after a steady flow. Red circles:  measured response in shear stress $\tau$ (left axis scale).  Dashed line: exponential fit to $\tau$ to obtain the relaxation timescale $T_1$.  (b) Blue squares: measured normal stress (right axis scale).  This relaxation behavior is qualitatively consistent with the generalized Newtonian model.
 }
\label{fig:stress_time_lowphi}
\end{figure}

%example rate-control, low phi
Figure \ref{fig:stress_time_lowphi}a shows  an example of  both the controlled shear rate $\dot\gamma$ and the response in shear stress $\tau$ as a function of time, for a sample at low $\phi_{eff}=58.1$ \% ($\dot\gamma_c^{-1} = 0.24$ s).  After the initial steady behavior for $t<0$ (initially at $\approx100$ Pa), there was a transient of the tool around $t=0$  before the shear rate settled down to the set point for $t>0$.  We note that the rheometer responds more slowly in rate-control than stress control, so the transient time for the shear rate to change from its initial setpoint value to zero is typically much  longer than for the stress in Figs.~\ref{fig:watercalibration}, \ref{fig:shearrate_time_newtonian}, and \ref{fig:shearrate_time_nonnewtonian}. 
 Data in this range is significantly affected by the PID control loop that controls the rotation of the tool, and should not be considered part of the sample response.
  
%material response - low phi
After this transient of the tool,  we observe a large negative stress (peak recorded magnitude of $-260$ Pa) in Fig.~\ref{fig:stress_time_lowphi}a.   This negative stress is to be expected in a rate-controlled flow even for a Newtonian fluid: after the plate stops but the suspension continues to flow, it applies a drag force on the top plate pushing it in its initial direction of motion, but with a force in the opposite direction as when the plate was driving the flow.  

As the top plate decelerates in rate-controlled experiments, the mean flow profile for a Newtonian fluid would have to transition from  a  plane Couette flow in the  steady state for $t<0$  to a more  parabolic profile that satisfies the no-slip condition at the top plate after it stopped moving.  This is  in contrast to stress control experiments where the mean flow profile for a Newtonian fluid could remain that of a plane Couette flow during the entire  experiment.   This change in profile requires a rapid energy dissipation as the shear rate near the top plate becomes very large during the transient compared to the steady-state profile, and a corresponding large torque on the tool during the transient.  Once the shear profile evolves from its initial plane Couette profile to a more parabolic profile, the drag force on the plate is expected  again to oppose the initial direction of flow, resulting in a positive stress.  After this transient with the negative stress,  we observed a decay in stress from a positive value.   We find the  stress at the start of this slower decay is $1.5\tau_{min}$, analogous to the stress-controlled experiments  in  Fig.~\ref{fig:shearrate_time_newtonian}. It appears that the faster relaxation  at higher effective viscosity in the shear thickening regime occurred over about the same time interval as the  flow-profile evolution.   Despite the  unusual transient behavior immediately after $t=0$, the relaxation behavior at this weight  fraction appears to be consistent with the generalized Newtonian model. 

In rate-controlled experiments, once the transient of the tool and the shear profile evolution are done, the expected solution for the shear stress decay for a constant viscosity, i.e.~where $\tau\propto \dot\gamma$, is the same as Eq.~\ref{eqn:Tr1_stresscontrol} with a swap of $\tau$ for $\dot\gamma$.  To obtain a relaxation timescale $T_1$ for rate-controlled experiments, we fit the measured $\tau(t)$  during the relaxation to the  exponentially decaying function
\begin{equation}
\tau \propto \exp(-t/T_1) + constant \ .
\label{eqn:Tr1}
\end{equation}
The addition of the constant accounts for the yield stress found at higher weight fractions. An example fit is shown in Fig.~\ref{fig:stress_time_lowphi}a. To avoid a contribution from the  transient of the tool immediately after the shear rate was set to zero, we generally started fitting after $\dot\gamma$  was less than 5\% of its initial $t<0$ set point value, and in cases like in Fig.~\ref{fig:stress_time_lowphi} where the sign of the shear stress first went negative after $t=0$, we fit only after the stress was positive again.   The  input error on fits was adjusted until the reduced $\chi^2 \approx 1$ to obtain an error on the fit value of $T_1$, as in the fits of Eq.~\ref{eqn:Tr1_stresscontrol}. The data in Fig.~\ref{fig:stress_time_lowphi} were fit with no  constant term, resulting in a good fit of the predicted exponential relaxation, which is qualitatively consistent with the generalize Newtonian model, even though there is a large scatter in which the root-mean-square difference  between the data and the fit over the first 0.6 s of the relaxation is 17\% of the proportionality coefficient. 
%error of 0.23, amplitude of 1.38, tau_1=0.93\pm0.01 when fit over 0.6 s.     
 
% Normal stress
The  measured normal stress for the rate-controlled relaxation experiment at $\phi_{eff}=58.1\%$ is shown in Fig.~\ref{fig:stress_time_lowphi}b.   In  this case the normal stress was negative for $t>0$, and relaxed along with the shear stress,  analogous to what was found  in stress-controlled measurements in the same weight fraction range in Fig.~\ref{fig:shearrate_time_newtonian}.

 \begin{figure}
\centering
\includegraphics[width=0.475\textwidth]{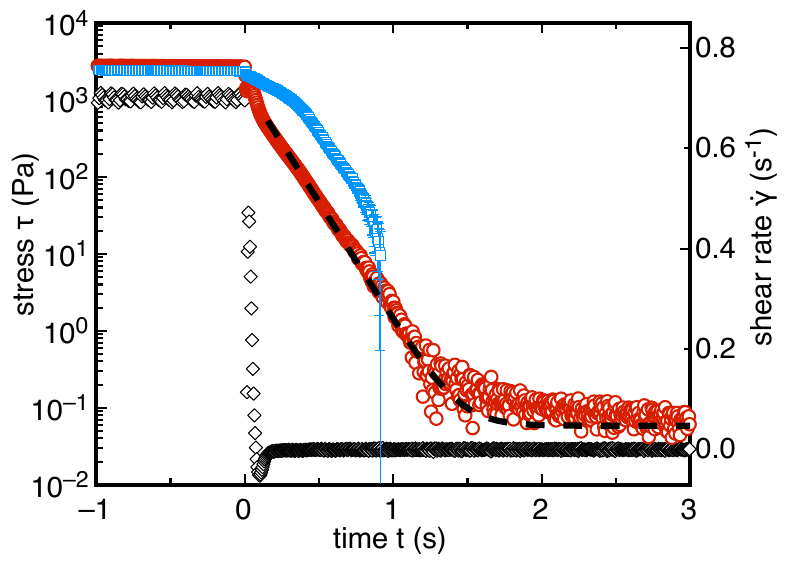}
\caption{(color online) An example of a transient rate-controlled experiment on the low side of the high-weight-fraction range ($\phi_{eff}=59.7\%$). Black diamonds:  the controlled shear rate $\dot\gamma$ (right axis scale).  Red circles:  measured shear stress $\tau$ response (left axis scale). Blue squares: measured normal stress $\tau_N$.  Dashed line:  exponential fit to $\tau$ to obtain the relaxation timescale $T_1$.  The  lack of a quick drop in shear stress to negative values or near $\tau_{min}$ just after $t=0$ is  inconsistent with the generalized Newtonian model.
}
\label{fig:stress_time_single}
\end{figure}
 
 %example rate-control, moderately high phi ( inconsistent with generalized Newtonian), single relaxation
 Figure \ref{fig:stress_time_single} shows  an example of the stress relaxation for a sample on the low side of the high-weight-fraction range, $\phi_{eff}=59.7$ \% ($\dot\gamma_{c}^{-1} = 2.4$ s).
% ($\phi_{eff}/\phi_c=0.979$)  
 %\gamma}_{c}^{-1} = 2.4$ s  
 The lack of an immediate drop to a negative shear stress or to near $\tau_{min}$ after $t=0$ is  qualitatively different from the stress-controlled relaxation observed in  the same weight fraction range in Fig.~\ref{fig:shearrate_time_nonnewtonian}.   Despite this observation -- which is qualitatively inconsistent with the generalized Newtonian model -- the exponential fit of Eq.~\ref{eqn:Tr1} to obtain $T_1$ is very good, with a root-mean-square difference of 2.4\% of the initial stress, indicating that we can still obtain a clear good measure of energy dissipation.

 %normal stress
 The  measured normal stress is also shown in Fig.~\ref{fig:stress_time_single}.  We only show the normal stress down to the measurement resolution ($\sim 10$ Pa), and plot this systematic error as the error bar in the figure.  In  this case the normal stress was positive,  similar to that found  in stress-controlled measurements in the same weight fraction range in Fig.~\ref{fig:shearrate_time_nonnewtonian}.  It tends to follow a similar trend as the shear stress, however, it is not proportional to the shear stress over a wide range of stress as is typical of steady-state DST \cite{LDHH05, BJ12}. 
 
\begin{figure}
\centering
\includegraphics[width=0.475\textwidth]{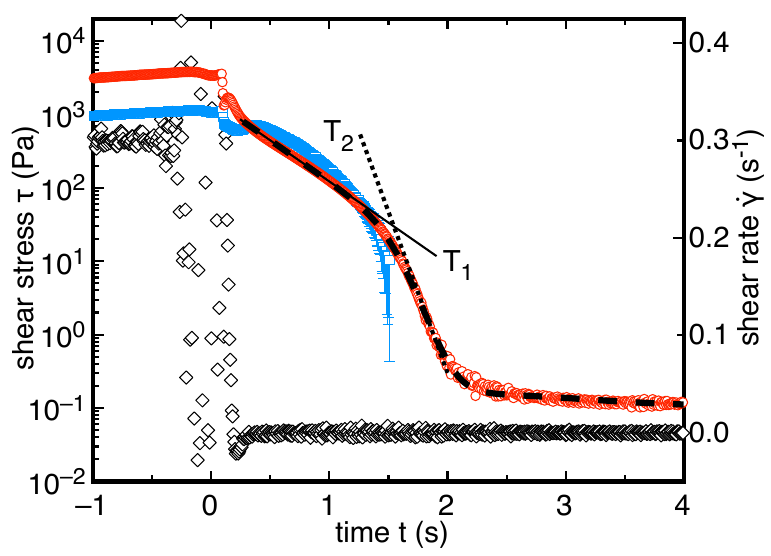}
\caption{(color online) An example of a transient rate-controlled experiment on the high side of the  high-weight-fraction range ($\phi_{eff} = 60.1$ \%), where two  exponential scaling regimes are found.  Black open diamonds:  the controlled shear rate $\dot\gamma$ (right axis scale).  Red open circles:  measured shear stress $\tau$ response (left axis scale). Dashed line:  exponential fit of Eq.~\ref{eqn:Tr1_Tr2}  to obtain the relaxation timescales $T_1$ (dotted line) and $T_2$ (solid line). Blue squares: measured normal stress $\tau_N$. The  lack of a quick drop in shear stress to negative values or near $\tau_{min}$ just after $t=0$ is  inconsistent with the generalized Newtonian model.
} 
\label{fig:stress_time_dual}
\end{figure}

%double-relaxation
At the highest weight fractions below $\phi_c$, the stress relaxation appears to have two exponential regimes, as shown for example in  Fig.~\ref{fig:stress_time_dual}  at $\phi_{eff} = 60.1$ \% ($\dot{\gamma}_{c}^{-1} = 9.1$ s).   Other than the two  exponential scaling regimes, the relaxation  behavior is qualitatively similar to Fig.~\ref{fig:stress_time_single}.  While the lack of a quick relaxation to $\tau_{min}$ or negative stress is inconsistent with the generalized Newtonian model, the two exponential ranges could in principle correspond to constant viscosity regions of a steady-state viscosity curve.  To test this, we obtain relaxation times $T_1$ and $T_2$ for each regime, using the fit function
\begin{equation}
\tau \propto \left( \frac{1}{\exp(-t/T_{1})} + \frac{1}{\exp(-t/T_{2})} \right)^{-1} + constant \ .
\label{eqn:Tr1_Tr2}
\end{equation}
We fit this to the data in Fig.~\ref{fig:stress_time_dual}  in two steps to provide better fit stability.   The first step fits Eq.~\ref{eqn:Tr1} to the data  the same way as in Fig.~\ref{fig:stress_time_single}  to fit the earlier, slower relaxation.  After $T_1$  is determined from this fit,  we fit Eq.~\ref{eqn:Tr1_Tr2} to the data with only $T_2$ as a free parameter, and only to the range $\tau \le 10$ Pa, with the same error fitting technique.  

% Normal stress
The  measured normal stress is also shown in Fig.~\ref{fig:stress_time_dual}.  Again, we only show the normal stress down to the measurement resolution, and plot this systematic error in the figure.  In  this case the normal stress was again positive and followed a similar trend as the shear stress,  qualitatively similar to that found  in in Fig.~\ref{fig:stress_time_single}.

\subsection{Weight fraction regimes of different rate-controlled relaxation behaviors}
\label{sec:phi_ranges_rate_control}

\begin{figure}
\centering
\includegraphics[width=0.475\textwidth]{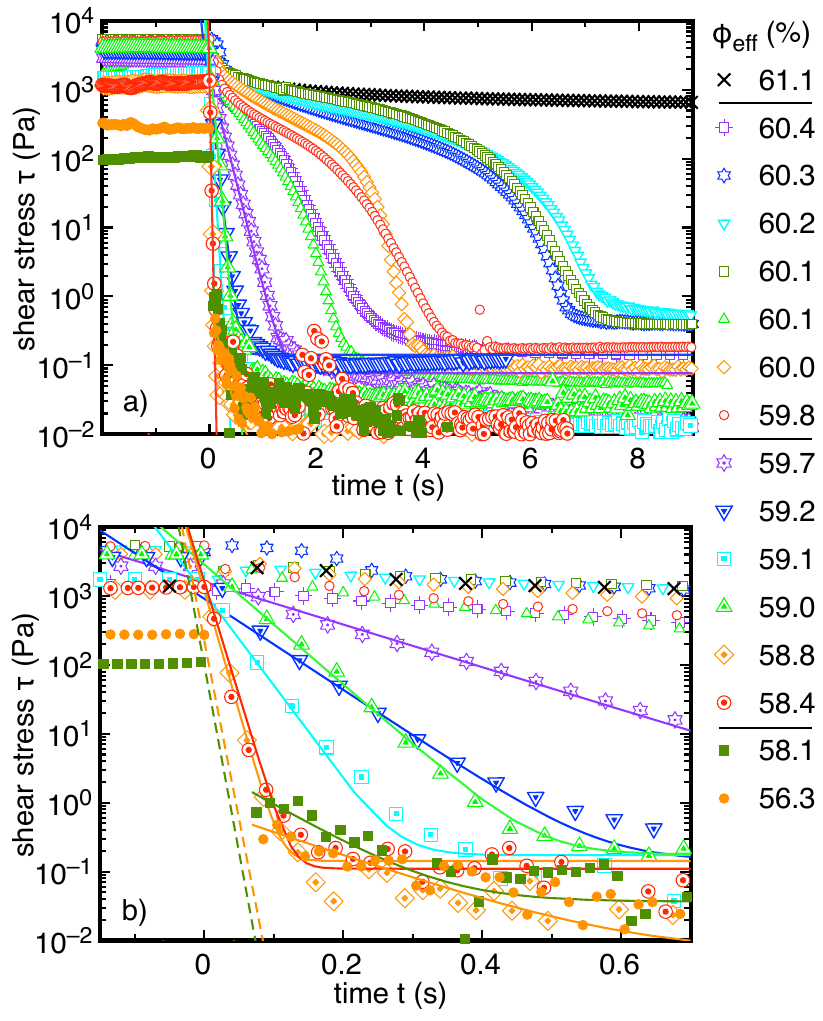}
\caption{(color online) Relaxation of shear stress $\tau$ over time $t$ for different weight fractions $\phi_{eff}$ listed in the legend. For ease of comparison, the symbols are the same as used in Fig.~\ref{fig:stress_shearrate_all}, and different filling types correspond to different qualitative behaviors.  (a) Dual-exponential  relaxation  is found for $59.8\% \le \phi_{eff} < 61.0\%$ (open symbols), while single  exponential relaxation is found for $58.4\% \le \phi_{eff} \le 59.7\%$ (partially filled symbols). (b) Zoomed scale of panel a to show that  for $\phi_{eff} \le 58.1\%$ (closed symbolss), there  is a faster initial relaxation followed by an exponential decay. Solid lines: fits of Eq.~\ref{eqn:Tr1} to the samples with single exponential behavior. Dashed lines: upper limit of the fast, early relaxation behavior for $\phi_{eff} \le 58.1\%$ based on the generalized Newtonian model of Eq.~\ref{eqn:Trmodel} using the maximum viscosity of the shear thickening range $\eta_{max}$,  confirming the fast relaxation is consistent with the generalized Newtonian model. 
} 
\label{fig:stress_time_all}
\end{figure}

%comparison at different  weight fractions - panel a
To determine the range of weight fraction ranges of the different relaxation behaviors seen in Figs.~\ref{fig:stress_time_lowphi}-\ref{fig:stress_time_dual}, we show in Fig.~\ref{fig:stress_time_all}a examples of the stress relaxation $\tau(t)$ for rate-controlled data at all weight fractions measured.  We will use this to compare to the stress-controlled data in later sections.
%For ease of comparison, we use the same symbols as in Fig.~\ref{fig:stress_shearrate_all}, which shows data for the same samples as the rate-controlled relaxation experiments.  The weight fraction range of similar behaviors as described below is also indicated by the type of symbol used (i.e.~open, partially filled, or solid), and the type of symbol used matches that in Fig.~\ref{fig:shearrate_time_all} for ease of comparison to stress-controlled data. Furthermore, the legend is split by horizontal lines to divide these different types of relaxation behavior. The legend indicates the values of $\phi_{eff}$, except 

%dual exp
The dual-exponential relaxation behavior shown for example in Fig.~\ref{fig:stress_time_dual} is found throughout the range $59.8\% \le \phi_{eff} < \phi_c$ ($\dot\gamma_c^{-1} \ge 3.4$ s,  open symbols in Fig.~\ref{fig:stress_time_all}).  Many of these higher weight fraction samples relax to a non-zero  stress value in the limit of large time, consistent with the yield stress  measured from the steady-state measurements in Fig.~\ref{fig:stress_shearrate_all}.   The transition between the two  exponential scaling regimes was consistently found to be on the scale of  $\tau \sim 10^2$ Pa.  For comparison, in stress-controlled experiments the stress dropped almost instantaneously from the initial  steady-state value to 0, so there was no measurable time period  over which two different  exponential relaxation regimes in stress  could be observed in stress-controlled experiments if the stress range is what determines the relaxation rate.  Thus, the  observation of two exponential  relaxation regimes is not inconsistent with the  stress-controlled measurements in Fig.~\ref{fig:shearrate_time_all}.  

%single exponential
The single-exponential relaxation behavior shown for example in Fig.~\ref{fig:stress_time_single} is found in the range $58.4\% \le \phi_{eff} \le 59.7\%$ (0.33 s $\le \dot\gamma_c^{-1}\le 2.7$ s, partially filled open symbols in Fig.~\ref{fig:stress_time_all}).  These relax much faster than at higher $\phi_{eff}$.  None of the data for $\phi_{eff} \ge 58.4\%$ ($\dot\gamma_c^{-1} \ge 0.33$ s) show the initial drop to a negative stress or to near $\tau_{min}$ expected from the generalized Newtonian model. 

%  generalized Newtonian - low phi
To see the behavior at low weight fractions $\phi \le 58.1\%$ ($\dot\gamma_c^{-1} \le 0.24$ s, solid symbols) in Fig.~\ref{fig:stress_time_all}, we  show a zoomed in version  of panel a in panel b.  To better see the signal in the noisy background, the data shown here are smoothed over a range of 0.025 s for $\phi_{eff} \le 58.8\%$ and a range of 0.050 s for other data sets after the shear rate is set to zero.   For $\phi \le 58.1\%$, extrapolations of the exponential fits of Eq.~\ref{eqn:Tr1} to the raw data  fall well below the stress at $t<0$, but are on average $(1.6\pm0.9)\tau_{min}$, analogous to the  stress-controlled experiments in the low-weight-fraction range shown in Fig.~\ref{fig:shearrate_time_all}, suggesting the  initial fast relaxation is again due to the higher viscosity for $\tau>\tau_{min}$.

We  note that we performed rate-controlled relaxation experiments at 5 different weight fractions in the range $56.0\% < \phi_{eff}\le 58.1\%$, but  in 3 of those cases we could not resolve any stress for $t>0$  as the stress was comparable to the noise level.  We do not present data for those cases here, or include it in later comparisons of $T_1$. 

%yield stress
For $\phi_{wt} =61.1\% > \phi_c$,  the  suspension behaves as a yield stress fluid as seen in Fig.~\ref{fig:stress_shearrate_all}, so there is no $\phi_{eff}$ scale and the raw value of $\phi_{wt}$ is given in Fig.~\ref{fig:stress_time_all} instead.   There is also no $\tau_{max}$, so in this case we instead set the initial shear rate based on the value of $\tau_{max}$ for $\phi_{eff}$ just below $\phi_c$.

\subsection{Calibration of transient viscosity for rate-controlled data}
\label{sec:calibration_rate_control}

Equation \ref{eqn:transientviscosity_calib} which relates the  relaxation time to transient viscosity is only valid for experiments performed under stress control, as it assumes the tool continued to rotate after we set $\tau = 0$. In shear-rate-controlled experiments, the tool was no longer rotating significantly at the relaxation set point $\dot\gamma = 0$. Therefore, the inertia of the tool no longer contributed to the angular momentum of the system, so $I_{tool}$ should be removed from Eq.~\ref{eqn:transientviscosity} when applied to rate-controlled flows. Furthermore, since the flow profile for a Newtonian fluid would transition from a  plane Couette flow in the  steady state for $t<0$  to a more  parabolic profile after the top plate stopped moving to satisfy the  no-slip boundary condition at the top plate in rate-controlled experiments, the  different flow profile would  result in a different proportionality coefficient for rate-controlled experiments than for stress-controlled experiments as given in Eq.~\ref{eqn:transientviscosity_calib}. Therefore, for rate-controlled experiments, we reduce the model from Eq.~\ref{eqn:transientviscosity_calib} to 
\begin{equation}
\eta_t \propto \dfrac{\rho d^2}{T_1} \ .
\label{eqn:transientviscosity_ratecontrol_scaling}
\end{equation}

%inability to calibrate directly
We attempted a calibration under rate-control with water.  Initially we observed a large negative stress, similar to the example in the generalized Newtonian regime in Fig.~\ref{fig:stress_time_lowphi}.    For $t>0$, large fluctuations in the measured stress drowned out any  relaxation signal we  attempted to measure.
%During this transient, we observed large negative stresses (up to 8000 Pa for water), before it increased back up. 
Even without the contribution of $I_{tool}$ to the total inertia of the system, the relaxation time would be expected to be resolvable based on Eq.~\ref{eqn:transientviscosity_ratecontrol_scaling}. On the other hand, this lack of measured stress is consistent with the data for $\phi_{eff} \le 58.1\%$ in Fig.~\ref{fig:stress_time_all}  where the stress dropped by 1-3  orders of magnitude during the tool deceleration and evolution of the shear profile, which would put the stress  below our measurement resolution for water.  While we were  unable to calibrate the rate-controlled experiments directly with this technique, this observation confirms that the data for $\phi_{eff} \le 58.1\%$ in Fig.~\ref{fig:stress_time_all} are consistent with the expectations of a generalized Newtonian fluid.

%fitting calibration constant
Instead, we  calibrate the  relaxation times from rate-controlled experiments by fitting to results from stress-controlled data.  The prediction  of the relaxation time from the generalized Newtonian model (Eq.~\ref{eqn:transientviscosity_calib}) agrees with the stress-controlled relaxation data at low weight fractions $\phi_{eff} \le 59.1$\%, seen quantitatively in Fig.~\ref{fig:viscosity_phi_phic},  as well as qualitatively,  as seen in Fig.~\ref{fig:shearrate_time_all}.  Likewise, the rate-controlled data for $\phi_{eff} \le 58.1\%$ were qualitatively consistent with the  generalized Newtonian model.   Thus, we can reasonably assume that the flow in rate-control in this weight fraction range is also quantitatively consistent with the generalized Newtonian model, obtain the scaling coefficient for Eq.~\ref{eqn:transientviscosity_ratecontrol_scaling}, and check for self-consistency later.  We obtain the scaling coefficient by  taking the ratio of the transient viscosity obtained from rate-controlled experiments to the steady-state viscosity from Eq.~\ref{eqn:transientviscosity_ratecontrol_scaling} assuming a scaling coefficient of 1,  and average over the 2 values of $\phi_{eff} \le 58.1\%$.  The scaling factor obtained is $15\pm7$. Hence, we can re-write Eq.~\ref{eqn:transientviscosity_ratecontrol_scaling} with a scaling factor as
% ($\phi_{eff}/\phi_c \le 0.968$)
\begin{equation}
\eta_t = 15 \dfrac{\rho d^2}{T_1}.
\label{eqn:transientviscosity_ratecontrol}
\end{equation}

%comparison to experiment -  generalized Newtonian behavior
The resulting values of  the transient viscosity $\eta_t$ for both $T_1$ and $T_2$ are shown in Fig.~\ref{fig:viscosity_phi_phic}.  The  values of $\eta_t$ from rate-controlled data using this scaling are consistent with the steady-state $\eta_{min}$ within their errors of 30\% for the same samples for $\phi_{eff}\le 58.8\%$.  These $\eta_t$  values are consistent with the generalized Newtonian model over this wider range than the fit.  %In the same range of $\phi_{eff}$, these values of $\eta_t$ also follow the same trends and are within the generous sample-to-sample scatter of the values of $\eta_t$  obtained from stress-controlled experiments (Eq.~\ref{eqn:transientviscosity_ratecontrol_scaling}), confirming the consistency of relaxation behavior in both stress- and rate-controlled experiments in this range.
 
%. self consistency check -  upper bound of  initial relaxation
As another self-consistency check for the calibration of  Eq.~\ref{eqn:transientviscosity_ratecontrol}, we use it to estimate the timescale of the initial fast relaxation for the data for $\phi\le 58.1\%$ shown in Fig.~\ref{fig:stress_time_all}b.  We plot  the shear stress as the dashed lines in Fig.~\ref{fig:stress_time_all}b for the predicted exponential decay of the generalized Newtonian model from Eq.~\ref{eqn:Tr1} using the maximum viscosity of the shear thickening range $\eta_{max}$ as an estimate of the transient viscosity, since the effective viscosity is close to $\eta_{max}$ in the steady-state for the initial shear rates just above $\dot\gamma_c$.  This is a rough  estimate because the shear profile must change in rate-controlled relaxation experiments during this fast decay, which  could change the dissipation rate relative to that of the steady-state shear profile. The fact that this prediction is about a factor of 2 faster than the measured relaxation to near $\tau_{min}$, where the  exponential decay starts is again plausibly consistent with the generalized Newtonian model, as far as we can resolve in this  weight fraction range.

%deviation
The transient viscosities $\eta_t$ do start deviating significantly from the steady state $\eta_{min}$ for $\phi_{eff} \ge 59.0\%$ ($\dot\gamma_c^{-1} \ge 0.60$ s), in the middle of the range  where the single-exponential relaxation behavior shown for example in Fig.~\ref{fig:stress_time_single} was found.  This deviation reaches up to 4 orders of magnitude at the largest $\phi_{eff} < \phi_c$ that we measured.  In this range, the values of $\eta_t$ based on rate-controlled measurements disagree with those  based on stress-controlled measurements, by about a factor of 3 for $T_2$ and an  order of magnitude for $T_1$.  This disagreement in values of $\eta_t$ for different flows indicates that $\eta_t$  is not an intrinsic property in this high $\phi_{eff}$ range.

\subsection{Comparison of time scales}
\label{sec:timescales}

\begin{figure}
\centering
\includegraphics[width=0.475\textwidth]{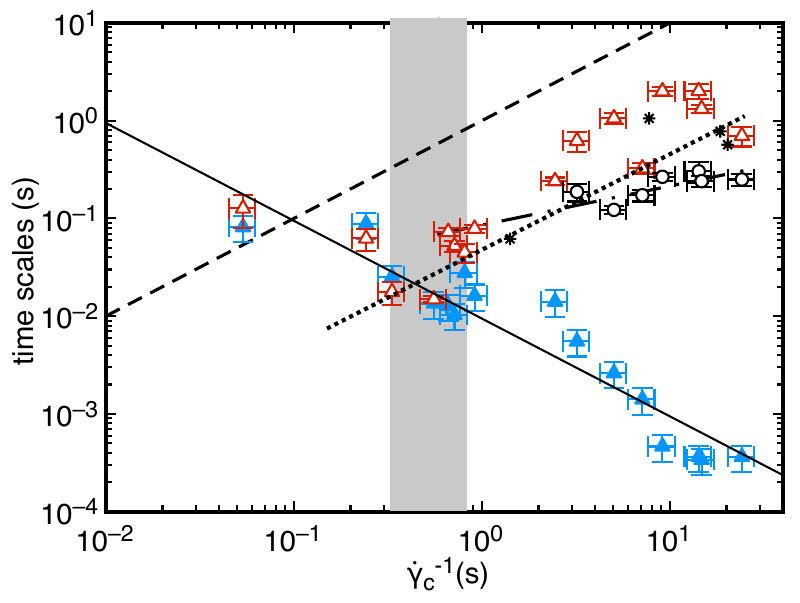}
\caption{(color online) Relaxation times as a function of  inverse critical shear rate  $\dot\gamma_c^{-1}$, which is a proxy for the weight fraction. %The top x-axis scale shows  some  corresponding values of $\phi_{eff}$ (not a linear scale) for reference.
Red open triangles:   $T_1$ for rate-controlled data.  Black open circles:  $T_2$ for rate-controlled data.  Blue solid triangles: generalized Newtonian model prediction $T_N$. Solid line: power law fit of $T_N$. Dotted line:  power law fit to $T_1$ in the range it deviates from the generalized Newtonian model.  Dashed-dotted line:  power law fit to $T_2$.  Dashed line:  the timescale corresponding to the  inverse critical shear rate $\dot{\gamma}_{c}^{-1}$, which  has a similar scaling as $T_1$ at large weight fraction. Black stars: $T_1$ for stress-controlled data, which agree with $T_1$ for rate-controlled data in the high-weight fraction range where the generalized Newtonian model fails (right side of the gray band), indicating the relaxation time is a more  universal quantity than viscosity in this range.
%The  measured relaxation timescales $T_1$ and $T_2$ both deviate from the generalized Newtonian model $T_N$ by several orders of magnitude and this difference diverges to infinity in the limit of large $\dot\gamma_c^{-1}$,  corresponding to $\phi\rightarrow \phi_c$ (the liquid-solid transition).   
 }
\label{fig:timescale_cricshearrate}
\end{figure}

%plot of  relaxation times vs. inverse critical shear rate
So far we have represented the relaxation behavior in terms of the transient viscosity $\eta_t$, but  we can make some different scaling comparisons,  and determine whether the viscosity or relaxation time is more intrinsic,  if we also plot the relaxation behavior in terms of timescales.  We show both relaxation times and the viscous dissipation timescale corresponding to the steady-state viscosity as a function of the time scale equal to the  inverse critical shear rate $\dot\gamma_c^{-1}$ in Fig.~\ref{fig:timescale_cricshearrate}.   $\dot\gamma_c^{-1}$ is also a proxy for the weight fraction, where $\phi_{eff}$ increases with $\dot\gamma_c^{-1}$, but with  much higher resolution near the critical point since it  diverges as $\phi_{eff}$  approaches $\phi_c$.  Rate-controlled data are replotted from  Fig.~\ref{fig:viscosity_phi_phic} to Fig.~\ref{fig:timescale_cricshearrate}.    A viscous dissipation timescale $T_N$ for the steady-state flow is  obtained from the steady-state viscosity $\eta_{min}$ based on Eq.~\ref{eqn:transientviscosity_ratecontrol} by replacing $\eta_t$ with $\eta_{min}$.  A power law fit to the corresponding timescale  yields an exponent -$0.92 \pm 0.05$, consistent within  2 standard deviations of the known inverse scaling between onset viscosity $\eta_{min} (\propto T_N^{-1})$ and  onset shear rate $\dot\gamma_c$ \cite{BJ09,FBOB12}.  The same general trends can be seen as in Fig.~\ref{fig:viscosity_phi_phic}, for example the transient time scale $T_1$ grows as $\dot\gamma_c^{-1}$ increases (i.e.~as  weight fraction increases) for $58.4\% \le\phi_{eff} \le \phi_c$ ($\dot\gamma_c^{-1} \ge 0.33$ s), while the viscous dissipation timescale $T_N$  based on the steady-state flow decreases.  

%bulk scaling
The viscous dissipation timescale predicted  from the generalized Newtonian model  is a quantity that depends on the dimensions and boundary conditions of the system, and is not an  intrinsic material property like the viscosity of a Newtonian fluid.  Equation \ref{eqn:Trmodel} predicts the viscous dissipation timescale $T_N$ coming from the  stress-controlled relaxation would be larger than  from the rate-controlled relaxation by a factor of $(I_{tool}+I)/I = 15$.  To test this, we plot the  relaxation time $T_1$  from stress-controlled data as stars in Fig.~\ref{fig:timescale_cricshearrate}, only plotting data in the range where the generalized Newtonian model fails for the stress-controlled data ($59.4\% \le \phi_{eff} < \phi_c$, $\dot\gamma_c^{-1} \ge 1.4$ s). We find that the stress-controlled relaxation times collapse with the rate-controlled $T_1$ in this range. This suggests that the relaxation time $T_1$ is  a more  universal quantity (i.e.~for  different types of flow control) than the viscosity  in  this high weight fraction range,  at least for a given system size.  Physically, it indicates that the relaxation time is not  determined by the time it takes to  dissipate the angular momentum of the system in this high weight fraction range according  to the generalized Newtonian model,  which is larger in stress-controlled experiments due to the addition of the tool inertia,  but could instead  be a  relaxation time of an internal structural of the suspension.
% I = 9.204e-7 kg m^2, I_tool = 1.282e-5 kg m^2, R = 0.025 m, $d=0.00125$ m

%power law fits and comparison to inverse of critical shear
%Percentage errors on all quantities are the same as used in Fig.~\ref{fig:viscosity_phi_phic}.
%were obtained from the fits of Eq.~\ref{eqn:Tr1} and \ref{eqn:Tr1_Tr2} by forcing the reduced $\chi^2\approx 1$. These fit errors were small compared to the standard deviation (typically 34\%) of multiple (typically 5) repetitions, so we plot only the standard deviation of the mean in Fig.~\ref{fig:timescale_cricshearrate}. The standard deviation of the mean for the repetitions of the relaxation times was on average 15\%.  
To gain some insight into the scaling of the relaxation time in the high weight fraction range, we least-squares fit a power law fit to $T_1(\dot\gamma_c^{-1})$ in the range where it deviates from the generalized Newtonian model in Fig.~\ref{fig:timescale_cricshearrate}.  We obtain a slope $1.0\pm 0.1$. We adjusted the input errors to the fit until the reduced $\chi^2 \approx 1$ so that the error represents an uncertainty on the slope assuming a power law fits the data.   An extrapolation of $T_1$ based on the power law fit of this range diverges to infinity in the limit of $\dot\gamma_c^{-1}\rightarrow \infty$ (dotted line),  corresponding to  the liquid-solid transition as $\phi \rightarrow \phi_c$ \cite{BJ09}.    A similar power law fit to $T_2$ in this range yields  an estimate for a slope of $ 0.4\pm 0.2$ (long dashed line).  For comparison with $T_1$, we plot the timescale $\dot\gamma_c^{-1}$ in Fig.~\ref{fig:timescale_cricshearrate}.  While this does not match any of the other time scales plotted,  the power law exponent is consistent with that of  $T_1$ of $1.0\pm0.1$ at large $\dot\gamma^{-1}$.  It is possible that   in the limit of large weight fraction, $T_1$ may be determined by $\dot\gamma_c$.  It was  proposed in some early shear thickening models  based on hydrodynamic mechanisms that $\dot\gamma_c^{-1}$ is proportional to a contact relaxation time -- a timescale it takes for particles in contact or near-contact to move significantly away from each other when they are pushed by the same repulsive forces that determine the onset of shear thickening, and resisted by viscous drag \cite{MB04a}.     The agreement with the scaling argument suggests that the unusually long  relaxation times we observe at high weight fractions may be due to this contact relaxation time.  
%When two separate  relaxation times are found for $59.8\% \le \phi_{eff} \le 61.0\%$ ($\dot\gamma_c^{-1} \ge 3.2$ s), they start to deviate from each other in Fig.~\ref{fig:timescale_cricshearrate}.

\section{Discussion}
\label{sec:discussion}

\subsection{Range of generalized Newtonian behavior}
% weight fraction ranges -  generalized Newtonian at low phi
In  both rate- and stress-controlled  relaxation experiments, we found suspensions to relax as a generalized Newtonian fluid where the  relaxation behavior can be described by the steady-state relation $\tau(\dot\gamma)$ at low weight fractions; specifically  $\phi_{eff}\le 59.1\%$ for stress-controlled data, and $\phi_{eff} \le 58.1\%$ in rate-controlled data.  We  found  the relaxation behavior to be qualitatively inconsistent with the generalized Newtonian model for $\phi_{eff}\ge 59.4\%$ for stress-controlled data, and $\phi_{eff} \ge 58.4\%$ for rate-controlled data.  Given the uncertainty of 0.3\% on of $\phi_{eff}$, we calculate a  best estimate of the transition as a mean of the four weight fractions as  $\phi_{eff} = 58.8\% \pm 0.3\%$ ($\dot\gamma_c^{-1}=0.54$ s.  This is consistent within about 1 standard deviation of the transitions measured separately from rate- or stress-controlled data.  This range is shown as the gray band in Figs.~\ref{fig:viscosity_phi_phic} and \ref{fig:timescale_cricshearrate}.   This gray band is also consistent with the intersection of the fits of $T_1$ and $T_N$ in Fig.~\ref{fig:timescale_cricshearrate} where the timescales start deviating from each other.  We henceforth refer to the weight fraction where the generalized Newtonian model starts to fail as  $\phi_{eff} = 58.8\% \pm 0.3\%$ when not referring to specific stress- or rate-controlled experiments.
%($\phi_{eff}/\phi_c=0.964\pm 0.011$)

 %transition from low to high phi
% For weight fractions  just above the  limit where we found relaxation  consistent with the generalized Newtonian model,  the  quantitative values of the transient viscosity in Fig.~\ref{fig:viscosity_phi_phic} were found to be close to the generalized Newtonian prediction for a narrow weight  fraction range [$\phi_{eff}\le59.4\%$ ($\dot\gamma_c^{-1}\le1.4$ s)  in stress-controlled measurements, and $\phi_{eff}\le58.8\%$ ($\dot\gamma_c^{-1}\le0.54$ s in rate-controlled measurements], despite the fact that the relaxation was qualitatively  inconsistent with generalized behavior.   This indicates a gradual transition in terms of the  relaxation time.  Extrapolation of the fit to $T_1$ in Fig.~\ref{fig:timescale_cricshearrate} crosses $T_N$ at the transition at $58.8\pm0.4\%$, confirming that this is a gradual transition, and suggesting the transition is a result of  whatever process is  responsible for the relaxation time $T_1$ becoming faster than the viscous dissipation timescale at higher weight fractions.
 
 %correspondence with viscosity curves at low phi
In the high-weight-fraction range ($58.8\% < \phi_{eff} < \phi_c$, $\dot\gamma_c^{-1} > 0.54$ s)  where the relaxation behavior was inconsistent with the generalized Newtonian model, we found  consistently discontinuous steady-state viscosity curves for the same samples shown in Fig.~\ref{fig:stress_shearrate_all}.  At  lower weight fractions,  whether or not the viscosity curves were  discontinuous was less consistent on repetition,  resulting in average curves that are steep, but  not discontinuous. From this correspondence we can conclude that the relaxation behavior is inconsistent with the generalized Newtonian model specifically at  weight fractions where the viscosity curve is consistently discontinuous in rate-controlled experiments.

\subsection{Normal stress and system-spanning structures} 
 
 %normal stress - low phi
 In the generalized Newtonian regime in the low weight fraction range ($\phi_{eff} < 58.8\pm 0.3\%$, $\dot\gamma_c^{-1} < 0.54$ s), we always observed negative normal stress during the relaxation. In contrast, for $\phi_{eff} > 58.8\pm0.3\%$   where we found disagreement with the generalized Newtonian model,  we always observed positive  normal stress  during the relaxation.   This correspondence   indicates the sign of the normal stress is also a good indicator of  whether or not the relaxation follows generalized Newtonian behavior.  
 
    % system spanning particle contact network mechanism for normal stress - high phi
In the high-weight-fraction regime  ($58.8\% < \phi_{eff} <\phi_c$, $\dot\gamma_c^{-1} > 0.54$ s), we observed a negative shear rate and oscillation during relaxation in stress control (e.g. Fig.~\ref{fig:shearrate_time_nonnewtonian}).   The positive normal stress observed during relaxation  could be the result of system-spanning  structures of particles pushing against each other.   The  positive normal stress in steady state DST  is often taken as  indication of a system-spanning network of particles in compression which push against each other and the rheometer plates \cite{BM97,BJ12, SMMD13}, and these networks are argued to play an important role in supporting frictional  interparticle contacts which transmit both normal and shear stresses \cite{BJ12, SMMD13}.   The plateau in stress at $\tau\approx\tau_{max}$ for higher shear rates in Fig.~\ref{fig:stress_shearrate_all} for this weight fraction range is consistent with the model of frustrated dilation \cite{BJ12},  where these normal and shear stresses are limited by  surface tension at the boundary, mostly independent of shear rate.  Thus, the interpretation of system-spanning structures of particles in contact at high weight fractions is consistent with steady-state interpretations.

  %hydrodynamic mechanisms for negative normal stress - low phi
In contrast,  the  negative normal stress  observed during relaxation in the generalized Newtonian regime ($\phi_{eff} < 58.8\%$, $\dot\gamma_c^{-1} < 0.54$ s) likely  originates from hydrodynamic effects.  This is also the  same weight fraction range where  we observed a linear relation in the steady-state $\tau(\dot\gamma)$  curve for $\tau>\tau_{max}$ (Fig.~\ref{fig:stress_shearrate_all}) for the same samples, instead of the plateau at $\tau\approx10^3$ Pa seen  at higher $\phi_{eff}$.  This feature is suggestive of a viscous hydrodynamic scaling  at the low weight fractions, which might be expected from the  hydrocluster model \cite{WB09}, or a model in which shear thickening is a transition between two viscous scaling regimes with different structure  \cite{WC14}. 

%normal stress positive in steady state and negative in transient -> short structural relaxation time
Perhaps surprisingly, the normal  stress was still positive  in the steady-state before the  transition to negative normal stress during relaxation in the generalized Newtonian regime.  The negative  normal stress during relaxation indicates that, if there was a system-spanning contact network  in the steady-state, its relaxation must have been fast compared to the viscous dissipation time $T_N$ in this weight fraction range.  The structural relaxation time corresponds to $T_1$ in the high-weight-fraction range in Fig.~\ref{fig:timescale_cricshearrate}, which follows the predicted scaling with $\dot\gamma_c^{-1}$ of the contact relaxation time it takes for particles  to separate against viscous drag \cite{MB04a}.  An extrapolation of the relaxation time $T_1$ in Fig.~\ref{fig:timescale_cricshearrate} drops below  the viscous dissipation time $T_N$ at a weight fraction consistent with that where the normal stress becomes negative during relaxation.  This confirms the structural relaxation time is shorter than the viscous dissipation time in the generalized Newtonian regime, and explains the transition from positive to negative normal stress during the transition from steady-state to relaxation.  The  intermittency of high stresses that makes discontinuous shear thickening intermittent  in the low-weight-fraction range for the steady-state viscosity curves of Fig.~\ref{fig:stress_shearrate_all} might also be the result of the structural relaxation time $T_1$ being comparable to the viscous dissipation time $T_N$. 
% too speculative? --  can we back up argument for negative normal stress with data/specific model?  If the steady-state shear develops compression  in a direction 45 degrees off from the direction of motion as is typical of simple shear, once the  applied stress or shear stops, then the  relaxation of the compressed structure would result in particles  moving away from each other and enlarging the lubrication gaps along the compression direction.  It  is possible that the negative normal stress observed is the result of this flow that enlarges the gaps.  
 
% explanation of lack of stress drop and positive shear stress  in rate controlled measurements
If the positive normal stress at high weight fractions ($58.8\% < \phi_{eff} <\phi_c$) is the result of a system spanning network of particle contacts,  most of the anomalous features of the shear stress relaxation in the rate-controlled experiments could also be explained.  The initially positive shear stress  during relaxation was inconsistent with the generalized Newtonian model (see Fig.~\ref{fig:stress_time_lowphi}), or any hydrodynamic model in which the suspension continued to flow during relaxation in the direction of initial motion of the plate.    If instead  a  temporary, nearly static structure forms after $t=0$, as in  the case of dynamic shear jamming \cite{PMJ16}, the built-up strain of the structure structure could  continue to apply  a stress on the plate opposite the original direction of motion, explaining the positive shear and normal stresses during rate-controlled relaxation.   Furthermore, if the local shear stress during relaxation is more dependent on normal stress via a solid-frictional coupling as in steady-state DST rather than on the local shear rate between particles \cite{BJ12}, then there is no longer  any expectation that the shear stress should  drop quickly to near $\tau_{min}$ according to the  steady-state $\tau(\dot\gamma)$ relation in the generalized Newtonian model, as seen in Fig.~\ref{fig:shearrate_time_nonnewtonian}.  Instead, the shear stress could remain on the scale of the normal stress, as seen in Figs.~\ref{fig:stress_time_single} and \ref{fig:stress_time_dual}.  

%structure and intrinsic relaxation time
Finally, the  observation that the relaxation time in the high-weight-fraction range ($58.8\% < \phi_{eff} <\phi_c$, $\dot\gamma_c^{-1} > 0.54$ s) is independent of the control mode (i.e.~stress-controlled or rate-controlled) is inconsistent  with the generalized Newtonian model in which the viscosity is intrinsic, or any model where the relaxation is limited by the need to dissipate the angular momentum of the tool, but it is expected if the  relaxation  is determined by the breakup of the system-spanning structure internal to the suspension.

\subsection{Oscillations}

%viscoelastic model
In the  high-weight-fraction range of $58.8\% < \phi_{eff} < \phi_c$, ($\dot\gamma_c^{-1} > 0.54$ s),  we found that the shear rate oscillates in stress-controlled relaxation  experiments,  as seen in Fig.~\ref{fig:shearrate_time_all}.  In particular the oscillation was found in cases where the damping was weak (i.e.~the relaxation time $T_1$ was large compared to the oscillation period).   This feature is indicative of  elastic energy storage in the system-spanning structure of particles in contact.  While the oscillations observed  in stress-controlled relaxation  experiments could be described by a viscoelastic model with shear modulus $G= 7$ Pa (Sec.~\ref{sec:stress_control}),  such a model would also predict oscillations when applied to rate-controlled measurements, with the only difference being whether or not the tool inertia is included.  However,  since we did not  observe oscillations in the rate control experiments in the same weight fraction range, we can rule out a straightforward viscoelastic extension to the generalized Newtonian model.  Therefore, another model is needed to describe these oscillations.

%oscillations and connection to  S-shaped curves, 
The oscillatory response  may also be related to previously observed S-shaped $\tau(\dot\gamma)$ curves in stress-controlled measurements \cite{NNM12, HGPPCW16}. In such $\tau(\dot\gamma)$ curves  there is an  unstable region of decreasing shear rate with increasing stress, which results in irregular oscillations from a low-stress liquid state to a high-stress solid-like state in steady-state stress-controlled measurements, but not in rate-controlled measurements  \cite{NNM12, HGPPCW16}.  Indeed,  the initial shear rates in the  stress-controlled relaxation measurements in Fig.~\ref{fig:shearrate_time_all} were much less than $\dot\gamma_c$,  even though we set the initial shear stress to be greater than $\tau_{max}$, only possible due to the S-shaped $\tau(\dot\gamma)$ curve.  We note that the other stress-controlled measurement  in the high weight fraction range at $\phi_{eff}=59.4\%$ showed no oscillation, but also had $\dot\gamma/\dot\gamma_c=21$ for $t<0$, which would put it  back on the stable branch of an S-shaped $\tau(\dot\gamma)$ curve, where no  oscillations would be expected.  Thus, the oscillatory behavior observed in our high-weight-fraction range ($58.8\% < \phi_{eff} < \phi_c$)  could be due to S-shaped $\tau(\dot\gamma)$ curves with a shear modulus $G$ for stress-controlled measurements only.   %While S-shaped $\tau(\dot\gamma)$ curves are sometimes seen over a wider  weight fraction range \cite{NNM12, HGPPCW16}, whether or not they are observed may depend sensitively on the control procedures.

\section{Conclusions}

% effective weight fraction
We showed that using the critical shear rate $\dot\gamma_c$ to characterize an effective weight fraction $\phi_{eff}$ can more precisely characterize material properties near the critical point at the liquid-solid transition $\phi_c$ with an uncertainty of 0.3\% (Fig.~\ref{fig:viscosity_phi}).  This high precision  allowed us to distinguish multiple transitions in behavior that are separated by about only 1\% in weight fraction,  something that could not have been done with larger uncertainties in weight fraction $\phi_{wt}$ directly measured by weight due to the suspensions' tendency to absorb  different amounts of water from the atmosphere at different temperature and humidity conditions. This conversion to $\phi_{eff}$ (Fig.~\ref{fig:phi_cricshearrate}) can also be used  to compare experiments done in other laboratories  or under different temperature and humidity conditions on a consistent $\phi_{eff}$ scale at our reference temperature  and humidity environment,  something which has not been achieved before without a measurement of $\phi_c$ due to the sensitivity of the weight fraction  of suspensions like cornstarch and water to temperature and humidity.  We caution the parameter values we obtained may still only apply for measurements at the same gap size, as the critical shear rate has been known to vary with gap size \cite{FBOB12}.

 %   comparison to generalized Newtonian model - low phi
Transient  measurements of stress relaxation over time revealed that DST fluids  exhibit  relaxation behavior consistent with a generalized Newtonian model in which the function $\tau(\dot\gamma)$  measured in steady state could describe transient  measurements of shear stress and shear rate during relaxation only for $\phi_{eff} < 58.8 \pm 0.3$ \%.  In this low weight-fraction range,  we found an initially quick relaxation -- though not well-resolved -- which was consistent with relaxation at viscosity $\eta_{max}$, approximately the viscosity of the initial steady-state before relaxation.  Once the shear rate or stress dropped below about the onset of shear thickening [$(2.2\pm0.5)\dot\gamma_c$ for  stress-controlled experiments  (Fig.~\ref{fig:shearrate_time_all}), or  $(1.6\pm0.9)\tau_{min}$ for rate-controlled experiments (Fig.~\ref{fig:stress_time_all})], we  observed an exponential relaxation with transient viscosity $\eta_t$ matching $\eta_{min}$,  approximately the viscosity at shear rates and stresses below the onset of shear thickening.  
 
% qualitative features of high phi range
However,  for $58.8\% < \phi_{eff} < \phi_c$ ($\dot\gamma_c^{-1} > 0.54$ s),  we found the relaxation behavior was inconsistent with the generalized Newtonian model.  We found the suspensions to  relax  without the  initial fast relaxation to $\tau_{min}$ in rate-controlled experiments, or $\dot\gamma_c$ in stress-controlled experiments, predicted by the generalized Newtonian model (Figs.~\ref{fig:shearrate_time_all}, \ref{fig:stress_time_all}).    In stress-controlled  measurements, we also observed  the shear rate initially became negative during relaxation, and oscillate in some cases.  These features can be described by a shear modulus $G$ that applies only in stress-controlled relaxation due to the history-dependence of S-shaped $\tau(\dot\gamma)$ curves \cite{NNM12, HGPPCW16}.  In rate-controlled experiments, for $59.8\% \le \phi_{eff} < \phi_c$ ($\dot\gamma_c^{-1} \ge 3.4$ s), we observed  two exponential  ranges could be fit to the shear stress relaxation in Fig.~\ref{fig:stress_time_all}a, in contrast to  the single exponential range observed elsewhere.    The scaling of  $T_1$ agrees with a prediction of the contact relaxation time for particles to separate \cite{MB04a}, while the physical origin of the second, faster relaxation time is unknown.

%   magnitude of viscosity discrepancy 
For $58.8\% < \phi_{eff} < \phi_c$, ($\dot\gamma_c^{-1} > 0.54$ s), the  transient viscosity values were found to decrease with $\phi_{eff}$, in contrast to the trend of  increasing  steady-state viscosity $\eta_{min}$ with $\phi_{eff}$.   The discrepancy was measured to be as large as 4 orders of magnitude. The extrapolated trends in Fig.~\ref{fig:viscosity_phi_phic} suggests the difference may diverge in the limit as $\phi_{eff} \rightarrow \phi_c$.  In this limit, the generalized Newtonian prediction approaches 0, while the fit of $T_1$ goes to infinity.  

%universality of relaxation time
 For $58.8\% < \phi_{eff} < \phi_c$ ($\dot\gamma_c^{-1} > 0.54$ s)  we also found that the relaxation time $T_1$ was a more consistent material property than viscosity when comparing stress- and rate-controlled measurements (Fig.~\ref{fig:timescale_cricshearrate}),  corresponding to a relaxation time of an internal structure. In contrast, in  the generalized Newtonian model, viscosity is the intrinsic property and relaxation time is  dependent on flow control.  Relaxation times are still expected to scale with system size in the high-weight-fraction regime, which was not investigated here.
 
%correspondence with transition in phi
We found a one-to-one correspondence between the weight fractions at which the steady-state viscosity curves were consistently discontinuous in rate-controlled measurements, with a nearly constant stress for $\tau>\tau_{max}$ (Fig.~\ref{fig:stress_shearrate_all}), and those at which the  relaxation was inconsistent with the generalized Newtonian behavior. This indicates a connection between the slow relaxation and discontinuous shear thickening.  We also found a one-to-one correspondence between positive normal stress during relaxation and inconsistency of the relaxation with the generalized Newtonian model (Figs.~\ref{fig:shearrate_time_nonnewtonian}, \ref{fig:stress_time_single}, \ref{fig:stress_time_dual}),  suggesting the continued existence of a system-spanning network of solid particle contacts during relaxation.  On the other hand, normal stresses which started out positive in the steady state became negative during relaxation at low weight fractions [$\phi_{eff} < 58.8\%$ ($\dot\gamma_c^{-1} < 0.54$ s)] may be a result of the structural relaxation time $T_1$ becomimg shorter than the viscous relaxation time $T_N$ (Fig.~\ref{fig:timescale_cricshearrate}). The persistence of this solid-like structure at high weight fractions  [$58.8\% < \phi_{eff} < \phi_c$ ($\dot\gamma_c^{-1} > 0.54$ s)]   accounts for many of the features in the relaxation behavior  that are inconsistent with the generalized Newtonian model, including the energy storage (Fig.~\ref{fig:shearrate_time_nonnewtonian}), the lack of initial fast relaxation while the shear stress remains coupled to the normal stress via frictional contacts (Figs.~\ref{fig:shearrate_time_newtonian} and \ref{fig:stress_time_lowphi}), and the lack of negative stress in rate-controlled relaxation due to the evolution of the shear profile (Fig.~\ref{fig:stress_time_lowphi}).

%Consequences of disagreement with the generalized Newtonian model
Disagreement with the generalized Newtonian model at high packing fractions indicates a failure in the assumptions used -- in particular the assumption that the steady-state $\tau(\dot\gamma)$ relation can fully describe the transient flow. While the solutions also assume a laminar flow, for a low-Reynolds-number flow like ours in a parallel plate rheometer, a non-laminar flow can only result in a mild discrepancy in which the reported steady-state $\eta_{min}$ could be an overestimate by as much as 1/3 of the yield stress, putting an upper bound on this correction of 33\%, which does not come close to the 4-orders-of-magnitude discrepancy we found with the generalized Newtonian model at the highest $\phi_{eff}$.  The failure of a $\tau(\dot\gamma)$ relation at the local level was previously known due to the dominance of a frictional term in the constitutive relation where shear stress is proportional to normal stress \cite{BJ12, FBOB12, FLO17}.  However, the dominance of the frictional term in $\tau(\dot\gamma)$ occurs over a much wider range of packing fractions, including the continuous shear thickening range \cite{BJ12, RBH16}, so the dominance of the frictional term in the constitutive relation does not in itself predict the failure of the generalized Newtonian model.  The failure to predict the relaxation time from the generalized Newtonian model leads to the surprising conclusion  that the relaxation is not controlled by the dissipative terms in the constitutive relation.  This discrepancy can be explained once the contact relaxation time is considered, which is larger than the dissipative relaxation time in the same packing fraction range  the generalized Newtonian model fails, so that the contact relaxation only slows the overall relaxation at the higher packing fractions.  Thus, to account for the full range of relaxation behavior we observed requires the addition of terms to the generalized Newtonian model including not only the frictional term that is needed for steady state, but also the shear modulus $G$, as well as the structural relaxation time $T_1$, and relaxation time of unknown origin $T_2$.  

 % comparison to glass transition
%The  deviation of the relaxation time from the viscous dissipation time scale approaching the liquid-solid transition is also reminiscent of a glass transition  \citep{EAN96}.  It has been established that  shear thickening suspensions do exhibit a jamming transition at this same weight fraction \cite{BJ09, BZFMBDJ11}, which  has long been assumed to be similar to a glass transition, specifically approached from the direction of increasing weight fraction \cite{LN98}.  Melting of a shear-jammed solid state to a liquid state under vibration has also been found to  have a diverging relaxation time  at the jamming transition in a system that exhibits shear thickening \cite{BBS02}.  At this time it is not obvious what the precise  connection to the  divergent relaxation times in these other systems is or if there is a connection. 

% application 
A nonzero relaxation time in the  limit of large weight fractions  may have  important consequences for the phenomena exhibited by DST fluids. For example, after an impacting object stops, if the  relaxation followed a generalized Newtonian model,  expected relaxation times would be less than 0.01 s in the discontinuous shear thickening range, where the response of cornstarch and water to impact is strongest \cite{MMASB17}.  This would be far too short for a  pool of cornstarch and water to  support a load like a solid long enough for a person  to step on it while they run across (a duration of typically 0.15 s) \cite{MMASB17,WSBW00}.  In contrast, we measured  relaxation times on the order of 1 s at the highest weight fractions below $\phi_c$, long enough to support a running person.  Other phenomena like the velocity oscillations  of a sinking sphere \cite{KSLM11} or rolling a sphere on the surface of the suspension \cite{youtube_bowlingball} would end much too fast to be observable by the naked eye based on the generalized Newtonian model.  For such phenomena to be  noticeable as dynamic with the naked eye requires a timescale on the order of seconds, which is in the range of what we find at large weight fractions.  How to specifically model such phenomena with a constitutive relation that includes a relaxation time, for example using the model of Ozgen et al. \cite{OBK15} is left open for future work.

 %In Newtonian liquids, the time scales seen are inversely proportional to the viscosity of the fluid. The bulk model of rheology predicts that the time scale present in the suspension is $T_r = \dfrac{1}{\dot{\gamma}_{c}}$ where $T_r$ is relaxation time scale (time taken by the system to relax), and $\dot{\gamma}_{c}$ is the critical shear rate --- the shear rate at which the system transitions from liquid to solid. Both of the models predict a single time scale in the fluid. However, the present study has found multiple relaxation time scales in the suspension sample. 

\section*{Acknowledgments}
We would like to thank Heinrich Jaeger for valuable discussions and suggestions. This work was supported by the National Science Foundation under grant DMR-1410157.

%
%\bibliography{../../rheology}

\begin{thebibliography}{36}%
\makeatletter
\providecommand \@ifxundefined [1]{%
 \@ifx{#1\undefined}
}%
\providecommand \@ifnum [1]{%
 \ifnum #1\expandafter \@firstoftwo
 \else \expandafter \@secondoftwo
 \fi
}%
\providecommand \@ifx [1]{%
 \ifx #1\expandafter \@firstoftwo
 \else \expandafter \@secondoftwo
 \fi
}%
\providecommand \natexlab [1]{#1}%
\providecommand \enquote  [1]{``#1''}%
\providecommand \bibnamefont  [1]{#1}%
\providecommand \bibfnamefont [1]{#1}%
\providecommand \citenamefont [1]{#1}%
\providecommand \href@noop [0]{\@secondoftwo}%
\providecommand \href [0]{\begingroup \@sanitize@url \@href}%
\providecommand \@href[1]{\@@startlink{#1}\@@href}%
\providecommand \@@href[1]{\endgroup#1\@@endlink}%
\providecommand \@sanitize@url [0]{\catcode `\\12\catcode `\$12\catcode
  `\&12\catcode `\#12\catcode `\^12\catcode `\_12\catcode `\%12\relax}%
\providecommand \@@startlink[1]{}%
\providecommand \@@endlink[0]{}%
\providecommand \url  [0]{\begingroup\@sanitize@url \@url }%
\providecommand \@url [1]{\endgroup\@href {#1}{\urlprefix }}%
\providecommand \urlprefix  [0]{URL }%
\providecommand \Eprint [0]{\href }%
\providecommand \doibase [0]{http://dx.doi.org/}%
\providecommand \selectlanguage [0]{\@gobble}%
\providecommand \bibinfo  [0]{\@secondoftwo}%
\providecommand \bibfield  [0]{\@secondoftwo}%
\providecommand \translation [1]{[#1]}%
\providecommand \BibitemOpen [0]{}%
\providecommand \bibitemStop [0]{}%
\providecommand \bibitemNoStop [0]{.\EOS\space}%
\providecommand \EOS [0]{\spacefactor3000\relax}%
\providecommand \BibitemShut  [1]{\csname bibitem#1\endcsname}%
\let\auto@bib@innerbib\@empty
%</preamble>
\bibitem [{\citenamefont {Lee}\ \emph {et~al.}(2003)\citenamefont {Lee},
  \citenamefont {Wetzel},\ and\ \citenamefont {Wagner}}]{LWW03}%
  \BibitemOpen
  \bibfield  {author} {\bibinfo {author} {\bibfnamefont {Y.~S.}\ \bibnamefont
  {Lee}}, \bibinfo {author} {\bibfnamefont {E.~D.}\ \bibnamefont {Wetzel}}, \
  and\ \bibinfo {author} {\bibfnamefont {N.J}\ \bibnamefont {Wagner}},\
  }\bibfield  {title} {\enquote {\bibinfo {title} {The ballistic impact
  characteristics of kevlar-woven fabrics impregnated with a colloidal shear
  thickening fluid},}\ }\href@noop {} {\bibfield  {journal} {\bibinfo
  {journal} {J. Materials Sci.}\ }\textbf {\bibinfo {volume} {38}},\ \bibinfo
  {pages} {2825} (\bibinfo {year} {2003})}\BibitemShut {NoStop}%
\bibitem [{\citenamefont {http://www.d3o.com/}()}]{D3O}%
  \BibitemOpen
  \bibfield  {author} {\bibinfo {author} {\bibnamefont {http://www.d3o.com/}},\
  }\href@noop {} {}\BibitemShut {NoStop}%
\bibitem [{\citenamefont {Barnes}(1989)}]{Ba89}%
  \BibitemOpen
  \bibfield  {author} {\bibinfo {author} {\bibfnamefont {H.~A.}\ \bibnamefont
  {Barnes}},\ }\bibfield  {title} {\enquote {\bibinfo {title} {Shear-thickening
  (``dilatancy'') in suspensions of nonaggregating solid particles dispersed in
  newtonian liquids},}\ }\href@noop {} {\bibfield  {journal} {\bibinfo
  {journal} {J. Rheology}\ }\textbf {\bibinfo {volume} {33}},\ \bibinfo {pages}
  {329} (\bibinfo {year} {1989})}\BibitemShut {NoStop}%
\bibitem [{you(2008)}]{youtube_running}%
  \BibitemOpen
  \href@noop {} {\enquote {\bibinfo {title}
  {https://www.youtube.com/watch?v=jks1ymq73oc,
  http://www.wimp.com/pool-filled-with-non-newtonian-fluid-cornstarch-and-water/},}\
  } (\bibinfo {year} {2008})\BibitemShut {NoStop}%
\bibitem [{\citenamefont {Brown}\ and\ \citenamefont {Jaeger}(2014)}]{BJ14}%
  \BibitemOpen
  \bibfield  {author} {\bibinfo {author} {\bibfnamefont {E.}~\bibnamefont
  {Brown}}\ and\ \bibinfo {author} {\bibfnamefont {H.~M.}\ \bibnamefont
  {Jaeger}},\ }\bibfield  {title} {\enquote {\bibinfo {title} {Shear thickening
  in concentrated suspensions: phenomenology, mechanisms, and relations to
  jamming},}\ }\href@noop {} {\bibfield  {journal} {\bibinfo  {journal}
  {Reports on Progress in Physics}\ }\textbf {\bibinfo {volume} {77}},\
  \bibinfo {pages} {046602--1--23} (\bibinfo {year} {2014})}\BibitemShut
  {NoStop}%
\bibitem [{\citenamefont {Merkt}\ \emph {et~al.}(2004)\citenamefont {Merkt},
  \citenamefont {Deegan}, \citenamefont {Goldman}, \citenamefont {Rericha},\
  and\ \citenamefont {Swinney}}]{MDGRS04}%
  \BibitemOpen
  \bibfield  {author} {\bibinfo {author} {\bibfnamefont {F.~S.}\ \bibnamefont
  {Merkt}}, \bibinfo {author} {\bibfnamefont {R.~D.}\ \bibnamefont {Deegan}},
  \bibinfo {author} {\bibfnamefont {D.~I.}\ \bibnamefont {Goldman}}, \bibinfo
  {author} {\bibfnamefont {E.~C.}\ \bibnamefont {Rericha}}, \ and\ \bibinfo
  {author} {\bibfnamefont {H.~L.}\ \bibnamefont {Swinney}},\ }\bibfield
  {title} {\enquote {\bibinfo {title} {Persistent holes in a fluid},}\
  }\href@noop {} {\bibfield  {journal} {\bibinfo  {journal} {Phys. Rev. Lett.}\
  }\textbf {\bibinfo {volume} {92}},\ \bibinfo {pages} {184501} (\bibinfo
  {year} {2004})}\BibitemShut {NoStop}%
\bibitem [{\citenamefont {von Kann}\ \emph {et~al.}(2011)\citenamefont {von
  Kann}, \citenamefont {Snoeijer}, \citenamefont {Lohse},\ and\ \citenamefont
  {van~der Meer}}]{KSLM11}%
  \BibitemOpen
  \bibfield  {author} {\bibinfo {author} {\bibfnamefont {S.}~\bibnamefont {von
  Kann}}, \bibinfo {author} {\bibfnamefont {J.~H.}\ \bibnamefont {Snoeijer}},
  \bibinfo {author} {\bibfnamefont {D.}~\bibnamefont {Lohse}}, \ and\ \bibinfo
  {author} {\bibfnamefont {D.}~\bibnamefont {van~der Meer}},\ }\bibfield
  {title} {\enquote {\bibinfo {title} {Non-monotonic settling of a sphere in a
  cornstarch suspension},}\ }\href@noop {} {\bibfield  {journal} {\bibinfo
  {journal} {Phys. Rev. E}\ }\textbf {\bibinfo {volume} {84}},\ \bibinfo
  {pages} {060401} (\bibinfo {year} {2011})}\BibitemShut {NoStop}%
\bibitem [{\citenamefont {Brown}\ and\ \citenamefont {Jaeger}(2009)}]{BJ09}%
  \BibitemOpen
  \bibfield  {author} {\bibinfo {author} {\bibfnamefont {E.}~\bibnamefont
  {Brown}}\ and\ \bibinfo {author} {\bibfnamefont {H.~M.}\ \bibnamefont
  {Jaeger}},\ }\bibfield  {title} {\enquote {\bibinfo {title} {Dynamic jamming
  point for shear thickening suspensions},}\ }\href@noop {} {\bibfield
  {journal} {\bibinfo  {journal} {Phys. Rev. Lett.}\ }\textbf {\bibinfo
  {volume} {103}},\ \bibinfo {pages} {086001} (\bibinfo {year}
  {2009})}\BibitemShut {NoStop}%
\bibitem [{\citenamefont {Brown}\ \emph {et~al.}(2011)\citenamefont {Brown},
  \citenamefont {Zhang}, \citenamefont {Forman}, \citenamefont {Maynor},
  \citenamefont {Betts}, \citenamefont {DeSimone},\ and\ \citenamefont
  {Jaeger}}]{BZFMBDJ11}%
  \BibitemOpen
  \bibfield  {author} {\bibinfo {author} {\bibfnamefont {E.}~\bibnamefont
  {Brown}}, \bibinfo {author} {\bibfnamefont {H.}~\bibnamefont {Zhang}},
  \bibinfo {author} {\bibfnamefont {N.~A.}\ \bibnamefont {Forman}}, \bibinfo
  {author} {\bibfnamefont {B.~W.}\ \bibnamefont {Maynor}}, \bibinfo {author}
  {\bibfnamefont {D.~E.}\ \bibnamefont {Betts}}, \bibinfo {author}
  {\bibfnamefont {J.~M.}\ \bibnamefont {DeSimone}}, \ and\ \bibinfo {author}
  {\bibfnamefont {H.~M.}\ \bibnamefont {Jaeger}},\ }\bibfield  {title}
  {\enquote {\bibinfo {title} {Shear thickening and jamming in densely packed
  suspensions of different particle shapes},}\ }\href@noop {} {\bibfield
  {journal} {\bibinfo  {journal} {Phys. Rev. E}\ }\textbf {\bibinfo {volume}
  {84}},\ \bibinfo {pages} {031408--1--11} (\bibinfo {year}
  {2011})}\BibitemShut {NoStop}%
\bibitem [{\citenamefont {Wagner}\ and\ \citenamefont {Brady}(2009)}]{WB09}%
  \BibitemOpen
  \bibfield  {author} {\bibinfo {author} {\bibfnamefont {N.J.}\ \bibnamefont
  {Wagner}}\ and\ \bibinfo {author} {\bibfnamefont {J.~F.}\ \bibnamefont
  {Brady}},\ }\bibfield  {title} {\enquote {\bibinfo {title} {Shear thickening
  in colloidal dispersions},}\ }\href@noop {} {\bibfield  {journal} {\bibinfo
  {journal} {Phys. Today, Oct. 2009}\ ,\ \bibinfo {pages} {27--32}} (\bibinfo
  {year} {2009})}\BibitemShut {NoStop}%
\bibitem [{\citenamefont {Waitukaitis}\ and\ \citenamefont
  {Jaeger}(2012)}]{WJ12}%
  \BibitemOpen
  \bibfield  {author} {\bibinfo {author} {\bibfnamefont {S.~R.}\ \bibnamefont
  {Waitukaitis}}\ and\ \bibinfo {author} {\bibfnamefont {H.~M.}\ \bibnamefont
  {Jaeger}},\ }\bibfield  {title} {\enquote {\bibinfo {title} {Impact-activated
  solidification of dense suspensions via dynamic jamming fronts},}\
  }\href@noop {} {\bibfield  {journal} {\bibinfo  {journal} {Nature}\ }\textbf
  {\bibinfo {volume} {487}},\ \bibinfo {pages} {205--209} (\bibinfo {year}
  {2012})}\BibitemShut {NoStop}%
\bibitem [{\citenamefont {Peters}\ and\ \citenamefont {Jaeger}(2014)}]{PJ14}%
  \BibitemOpen
  \bibfield  {author} {\bibinfo {author} {\bibfnamefont {I.~R.}\ \bibnamefont
  {Peters}}\ and\ \bibinfo {author} {\bibfnamefont {H.~M.}\ \bibnamefont
  {Jaeger}},\ }\bibfield  {title} {\enquote {\bibinfo {title} {Quasi-2d dynamic
  jamming in cornstarch suspensions: visualization and force measurements},}\
  }\href@noop {} {\bibfield  {journal} {\bibinfo  {journal} {Soft Matter}\
  }\textbf {\bibinfo {volume} {10}},\ \bibinfo {pages} {6574--6570} (\bibinfo
  {year} {2014})}\BibitemShut {NoStop}%
\bibitem [{\citenamefont {Maharjan}\ \emph {et~al.}(2017)\citenamefont
  {Maharjan}, \citenamefont {Mukhopadhyay}, \citenamefont {Allen},
  \citenamefont {Storz},\ and\ \citenamefont {Brown}}]{MMASB17}%
  \BibitemOpen
  \bibfield  {author} {\bibinfo {author} {\bibfnamefont {R.}~\bibnamefont
  {Maharjan}}, \bibinfo {author} {\bibfnamefont {S.}~\bibnamefont
  {Mukhopadhyay}}, \bibinfo {author} {\bibfnamefont {B.}~\bibnamefont {Allen}},
  \bibinfo {author} {\bibfnamefont {T.}~\bibnamefont {Storz}}, \ and\ \bibinfo
  {author} {\bibfnamefont {E.}~\bibnamefont {Brown}},\ }\bibfield  {title}
  {\enquote {\bibinfo {title} {Constitutive relation of the system-spanning
  dynamically jammed region in response to impact of cornstarch and water
  suspensions},}\ }\href@noop {} {\bibfield  {journal} {\bibinfo  {journal}
  {arXiv:1407.0719}\ } (\bibinfo {year} {2017})}\BibitemShut {NoStop}%
\bibitem [{\citenamefont {Deegan}(2010)}]{De10}%
  \BibitemOpen
  \bibfield  {author} {\bibinfo {author} {\bibfnamefont {R.~D.}\ \bibnamefont
  {Deegan}},\ }\bibfield  {title} {\enquote {\bibinfo {title} {Stress
  hysteresis as the cause of persistent holes in particulate suspensions},}\
  }\href@noop {} {\bibfield  {journal} {\bibinfo  {journal} {Phys. Rev. E}\
  }\textbf {\bibinfo {volume} {81}},\ \bibinfo {pages} {036319} (\bibinfo
  {year} {2010})}\BibitemShut {NoStop}%
\bibitem [{\citenamefont {von Kann}\ \emph {et~al.}(2013)\citenamefont {von
  Kann}, \citenamefont {Snoeijer},\ and\ \citenamefont {van~der Meer}}]{KSM13}%
  \BibitemOpen
  \bibfield  {author} {\bibinfo {author} {\bibfnamefont {S.}~\bibnamefont {von
  Kann}}, \bibinfo {author} {\bibfnamefont {J.~H.}\ \bibnamefont {Snoeijer}}, \
  and\ \bibinfo {author} {\bibfnamefont {D.}~\bibnamefont {van~der Meer}},\
  }\bibfield  {title} {\enquote {\bibinfo {title} {Velocity oscillations and
  stop-go cycles: The trajectory of an object settling in a cornstarch
  suspension},}\ }\href@noop {} {\bibfield  {journal} {\bibinfo  {journal}
  {Phys. Rev. E}\ }\textbf {\bibinfo {volume} {87}},\ \bibinfo {pages} {042301}
  (\bibinfo {year} {2013})}\BibitemShut {NoStop}%
\bibitem [{\citenamefont {Ozgen}\ \emph {et~al.}(2015)\citenamefont {Ozgen},
  \citenamefont {Brown},\ and\ \citenamefont {Kallman}}]{OBK15}%
  \BibitemOpen
  \bibfield  {author} {\bibinfo {author} {\bibfnamefont {O.}~\bibnamefont
  {Ozgen}}, \bibinfo {author} {\bibfnamefont {E.}~\bibnamefont {Brown}}, \ and\
  \bibinfo {author} {\bibfnamefont {M.}~\bibnamefont {Kallman}},\ }\bibfield
  {title} {\enquote {\bibinfo {title} {Simulating the dynamic behavior of shear
  thickening fluids},}\ }\href@noop {} {\bibfield  {journal} {\bibinfo
  {journal} {arXiv:1510.09069}\ } (\bibinfo {year} {2015})}\BibitemShut
  {NoStop}%
\bibitem [{you(2009)}]{youtube_bowlingball}%
  \BibitemOpen
  \href@noop {} {\enquote {\bibinfo {title}
  {https://www.youtube.com/watch?v=8seb0nhx5tu},}\ } (\bibinfo {year}
  {2009})\BibitemShut {NoStop}%
\bibitem [{\citenamefont {Lee}\ and\ \citenamefont {Wagner}(2003)}]{LW03}%
  \BibitemOpen
  \bibfield  {author} {\bibinfo {author} {\bibfnamefont {Y.~S.}\ \bibnamefont
  {Lee}}\ and\ \bibinfo {author} {\bibfnamefont {N.J}\ \bibnamefont {Wagner}},\
  }\bibfield  {title} {\enquote {\bibinfo {title} {Dynamic properties of shear
  thickening colloidal suspensions},}\ }\href@noop {} {\bibfield  {journal}
  {\bibinfo  {journal} {Rheol. Acta}\ }\textbf {\bibinfo {volume} {42}},\
  \bibinfo {pages} {199--208} (\bibinfo {year} {2003})}\BibitemShut {NoStop}%
\bibitem [{\citenamefont {Raghavan}\ and\ \citenamefont {Khan}(1997)}]{RK97}%
  \BibitemOpen
  \bibfield  {author} {\bibinfo {author} {\bibfnamefont {S.~R.}\ \bibnamefont
  {Raghavan}}\ and\ \bibinfo {author} {\bibfnamefont {S.~A.}\ \bibnamefont
  {Khan}},\ }\bibfield  {title} {\enquote {\bibinfo {title} {Shear-thickening
  response of fumed silica suspensions under steady and oscillatory shear},}\
  }\href@noop {} {\bibfield  {journal} {\bibinfo  {journal} {J. Colloid and
  Interface Science}\ }\textbf {\bibinfo {volume} {185}},\ \bibinfo {pages}
  {57--67} (\bibinfo {year} {1997})}\BibitemShut {NoStop}%
\bibitem [{\citenamefont {Brown}\ and\ \citenamefont {Jaeger}(2012)}]{BJ12}%
  \BibitemOpen
  \bibfield  {author} {\bibinfo {author} {\bibfnamefont {E.}~\bibnamefont
  {Brown}}\ and\ \bibinfo {author} {\bibfnamefont {H.}~\bibnamefont {Jaeger}},\
  }\bibfield  {title} {\enquote {\bibinfo {title} {The role of dilation and
  confining stress in shear thickening of dense suspensions},}\ }\href@noop {}
  {\bibfield  {journal} {\bibinfo  {journal} {J. Rheology}\ }\textbf {\bibinfo
  {volume} {56}},\ \bibinfo {pages} {875--923} (\bibinfo {year}
  {2012})}\BibitemShut {NoStop}%
\bibitem [{\citenamefont {Lin}\ \emph {et~al.}(2015)\citenamefont {Lin},
  \citenamefont {Guy}, \citenamefont {Hermes}, \citenamefont {Ness},
  \citenamefont {Sun}, \citenamefont {Poon},\ and\ \citenamefont
  {Cohen}}]{LGHNSPC15}%
  \BibitemOpen
  \bibfield  {author} {\bibinfo {author} {\bibfnamefont {Neil Y.~C.}\
  \bibnamefont {Lin}}, \bibinfo {author} {\bibfnamefont {Ben~M.}\ \bibnamefont
  {Guy}}, \bibinfo {author} {\bibfnamefont {Michiel}\ \bibnamefont {Hermes}},
  \bibinfo {author} {\bibfnamefont {Chris}\ \bibnamefont {Ness}}, \bibinfo
  {author} {\bibfnamefont {Jin}\ \bibnamefont {Sun}}, \bibinfo {author}
  {\bibfnamefont {Wilson C.~K.}\ \bibnamefont {Poon}}, \ and\ \bibinfo {author}
  {\bibfnamefont {Itai}\ \bibnamefont {Cohen}},\ }\bibfield  {title} {\enquote
  {\bibinfo {title} {Hydrodynamic and contact contributions to shear thickening
  in colloidal suspensions},}\ }\href@noop {} {\bibfield  {journal} {\bibinfo
  {journal} {Physical Review Letters}\ }\textbf {\bibinfo {volume} {115}},\
  \bibinfo {pages} {228304} (\bibinfo {year} {2015})}\BibitemShut {NoStop}%
\bibitem [{\citenamefont {Lootens}\ \emph {et~al.}(2003)\citenamefont
  {Lootens}, \citenamefont {Damme},\ and\ \citenamefont {H{\'e}braud}}]{LDH03}%
  \BibitemOpen
  \bibfield  {author} {\bibinfo {author} {\bibfnamefont {D.}~\bibnamefont
  {Lootens}}, \bibinfo {author} {\bibfnamefont {H.~Van}\ \bibnamefont {Damme}},
  \ and\ \bibinfo {author} {\bibfnamefont {P.}~\bibnamefont {H{\'e}braud}},\
  }\bibfield  {title} {\enquote {\bibinfo {title} {Giant stress fluctuations at
  the jamming transition},}\ }\href@noop {} {\bibfield  {journal} {\bibinfo
  {journal} {Phys. Rev. Lett.}\ }\textbf {\bibinfo {volume} {90}},\ \bibinfo
  {pages} {178301} (\bibinfo {year} {2003})}\BibitemShut {NoStop}%
\bibitem [{\citenamefont {Fall}\ \emph {et~al.}(2012)\citenamefont {Fall},
  \citenamefont {Bertrand}, \citenamefont {Ovarlez},\ and\ \citenamefont
  {Bonn}}]{FBOB12}%
  \BibitemOpen
  \bibfield  {author} {\bibinfo {author} {\bibfnamefont {A.}~\bibnamefont
  {Fall}}, \bibinfo {author} {\bibfnamefont {F.}~\bibnamefont {Bertrand}},
  \bibinfo {author} {\bibfnamefont {G.}~\bibnamefont {Ovarlez}}, \ and\
  \bibinfo {author} {\bibfnamefont {B.}~\bibnamefont {Bonn}},\ }\bibfield
  {title} {\enquote {\bibinfo {title} {Shear thickening of cornstarch
  suspensions},}\ }\href@noop {} {\bibfield  {journal} {\bibinfo  {journal} {J.
  Rheology}\ }\textbf {\bibinfo {volume} {56}},\ \bibinfo {pages} {575--591}
  (\bibinfo {year} {2012})}\BibitemShut {NoStop}%
\bibitem [{\citenamefont {Mewis}\ and\ \citenamefont {Wagner}(2012)}]{MJ12}%
  \BibitemOpen
  \bibfield  {author} {\bibinfo {author} {\bibfnamefont {Jan}\ \bibnamefont
  {Mewis}}\ and\ \bibinfo {author} {\bibfnamefont {Norman~J.}\ \bibnamefont
  {Wagner}},\ }\href@noop {} {\emph {\bibinfo {title} {Colloidal Suspension
  Rheology}}}\ (\bibinfo  {publisher} {Cambridge University Press},\ \bibinfo
  {year} {2012})\BibitemShut {NoStop}%
\bibitem [{\citenamefont {White}(2009)}]{Wh09}%
  \BibitemOpen
  \bibfield  {author} {\bibinfo {author} {\bibfnamefont {Frank~M.}\
  \bibnamefont {White}},\ }\href@noop {} {\emph {\bibinfo {title} {Fluid
  Mechanics}}},\ \bibinfo {edition} {7th}\ ed.\ (\bibinfo  {publisher}
  {McGraw-Hill},\ \bibinfo {year} {2009})\BibitemShut {NoStop}%
\bibitem [{\citenamefont {Lootens}\ \emph {et~al.}(2005)\citenamefont
  {Lootens}, \citenamefont {vanDamme}, \citenamefont {H{\'e}mar},\ and\
  \citenamefont {H{\'e}braud}}]{LDHH05}%
  \BibitemOpen
  \bibfield  {author} {\bibinfo {author} {\bibfnamefont {D.}~\bibnamefont
  {Lootens}}, \bibinfo {author} {\bibfnamefont {H.}~\bibnamefont {vanDamme}},
  \bibinfo {author} {\bibfnamefont {Y.}~\bibnamefont {H{\'e}mar}}, \ and\
  \bibinfo {author} {\bibfnamefont {P.}~\bibnamefont {H{\'e}braud}},\
  }\bibfield  {title} {\enquote {\bibinfo {title} {Dilatant flow of
  concentrated suspensions of rough particles},}\ }\href@noop {} {\bibfield
  {journal} {\bibinfo  {journal} {Phys. Rev. Lett.}\ }\textbf {\bibinfo
  {volume} {95}},\ \bibinfo {pages} {268302} (\bibinfo {year}
  {2005})}\BibitemShut {NoStop}%
\bibitem [{\citenamefont {Seto}\ \emph {et~al.}(2013)\citenamefont {Seto},
  \citenamefont {Mari}, \citenamefont {Morris},\ and\ \citenamefont
  {Denn}}]{SMMD13}%
  \BibitemOpen
  \bibfield  {author} {\bibinfo {author} {\bibfnamefont {R.}~\bibnamefont
  {Seto}}, \bibinfo {author} {\bibfnamefont {R.}~\bibnamefont {Mari}}, \bibinfo
  {author} {\bibfnamefont {J.~F.}\ \bibnamefont {Morris}}, \ and\ \bibinfo
  {author} {\bibfnamefont {M.~M.}\ \bibnamefont {Denn}},\ }\bibfield  {title}
  {\enquote {\bibinfo {title} {Discontinuous shear thickening of frictional
  hard-sphere suspensions},}\ }\href@noop {} {\bibfield  {journal} {\bibinfo
  {journal} {Phys.~Rev.~Lett.}\ }\textbf {\bibinfo {volume} {111}},\ \bibinfo
  {pages} {218301} (\bibinfo {year} {2013})}\BibitemShut {NoStop}%
\bibitem [{\citenamefont {Melrose}\ and\ \citenamefont {Ball}(2004)}]{MB04a}%
  \BibitemOpen
  \bibfield  {author} {\bibinfo {author} {\bibfnamefont {J.~R.}\ \bibnamefont
  {Melrose}}\ and\ \bibinfo {author} {\bibfnamefont {R.~C.}\ \bibnamefont
  {Ball}},\ }\bibfield  {title} {\enquote {\bibinfo {title} {Continuous shear
  thickening transitions in model concentrated colloids -- the role of
  interparticle forces},}\ }\href@noop {} {\bibfield  {journal} {\bibinfo
  {journal} {J. Rheology}\ }\textbf {\bibinfo {volume} {48}},\ \bibinfo {pages}
  {937} (\bibinfo {year} {2004})}\BibitemShut {NoStop}%
\bibitem [{\citenamefont {Brady}\ and\ \citenamefont {Morris}(1997)}]{BM97}%
  \BibitemOpen
  \bibfield  {author} {\bibinfo {author} {\bibfnamefont {J.~F.}\ \bibnamefont
  {Brady}}\ and\ \bibinfo {author} {\bibfnamefont {J.~F.}\ \bibnamefont
  {Morris}},\ }\bibfield  {title} {\enquote {\bibinfo {title} {Microstructure
  of strongly sheared suspensions and its impact on rheology and diffusion},}\
  }\href@noop {} {\bibfield  {journal} {\bibinfo  {journal} {J. Fluid Mech.}\
  }\textbf {\bibinfo {volume} {348}},\ \bibinfo {pages} {103--139} (\bibinfo
  {year} {1997})}\BibitemShut {NoStop}%
\bibitem [{\citenamefont {Wyart}\ and\ \citenamefont {Cates}(2014)}]{WC14}%
  \BibitemOpen
  \bibfield  {author} {\bibinfo {author} {\bibfnamefont {M.}~\bibnamefont
  {Wyart}}\ and\ \bibinfo {author} {\bibfnamefont {M.~E.}\ \bibnamefont
  {Cates}},\ }\bibfield  {title} {\enquote {\bibinfo {title} {Discontinuous
  shear thickening without inertia in dense non-brownian suspensions},}\
  }\href@noop {} {\bibfield  {journal} {\bibinfo  {journal} {Phys. Rev. Lett.}\
  }\textbf {\bibinfo {volume} {112}},\ \bibinfo {pages} {098302} (\bibinfo
  {year} {2014})}\BibitemShut {NoStop}%
\bibitem [{\citenamefont {Peters}\ \emph {et~al.}(2016)\citenamefont {Peters},
  \citenamefont {Majumdar},\ and\ \citenamefont {Jaeger}}]{PMJ16}%
  \BibitemOpen
  \bibfield  {author} {\bibinfo {author} {\bibfnamefont {I.~R.}\ \bibnamefont
  {Peters}}, \bibinfo {author} {\bibfnamefont {S.}~\bibnamefont {Majumdar}}, \
  and\ \bibinfo {author} {\bibfnamefont {H.~M.}\ \bibnamefont {Jaeger}},\
  }\bibfield  {title} {\enquote {\bibinfo {title} {Direct observation of
  dynamic shear jamming in dense suspensions},}\ }\href@noop {} {\bibfield
  {journal} {\bibinfo  {journal} {Nature}\ }\textbf {\bibinfo {volume} {532}},\
  \bibinfo {pages} {214--217} (\bibinfo {year} {2016})}\BibitemShut {NoStop}%
\bibitem [{\citenamefont {Nakanishi}\ \emph {et~al.}(2012)\citenamefont
  {Nakanishi}, \citenamefont {Nagahiro},\ and\ \citenamefont
  {Mitarai}}]{NNM12}%
  \BibitemOpen
  \bibfield  {author} {\bibinfo {author} {\bibfnamefont {Hiizu}\ \bibnamefont
  {Nakanishi}}, \bibinfo {author} {\bibfnamefont {S.}~\bibnamefont {Nagahiro}},
  \ and\ \bibinfo {author} {\bibfnamefont {Namiko}\ \bibnamefont {Mitarai}},\
  }\bibfield  {title} {\enquote {\bibinfo {title} {Fluid dynamics of dilatant
  fluids},}\ }\href@noop {} {\bibfield  {journal} {\bibinfo  {journal} {Phys.
  Rev. E}\ }\textbf {\bibinfo {volume} {85}},\ \bibinfo {pages} {011401}
  (\bibinfo {year} {2012})}\BibitemShut {NoStop}%
\bibitem [{\citenamefont {Hermes}\ \emph {et~al.}(2016)\citenamefont {Hermes},
  \citenamefont {Guy}, \citenamefont {Poon}, \citenamefont {Poy}, \citenamefont
  {Cates},\ and\ \citenamefont {Wyart}}]{HGPPCW16}%
  \BibitemOpen
  \bibfield  {author} {\bibinfo {author} {\bibfnamefont {Michiel}\ \bibnamefont
  {Hermes}}, \bibinfo {author} {\bibfnamefont {Ben~M.}\ \bibnamefont {Guy}},
  \bibinfo {author} {\bibfnamefont {Wilson C.~K.}\ \bibnamefont {Poon}},
  \bibinfo {author} {\bibfnamefont {Guilhem}\ \bibnamefont {Poy}}, \bibinfo
  {author} {\bibfnamefont {Michael~E.}\ \bibnamefont {Cates}}, \ and\ \bibinfo
  {author} {\bibfnamefont {Matthieu}\ \bibnamefont {Wyart}},\ }\bibfield
  {title} {\enquote {\bibinfo {title} {Unsteady flow and particle migration in
  dense, non-brownian suspensions},}\ }\href@noop {} {\bibfield  {journal}
  {\bibinfo  {journal} {Journal of Rheology}\ }\textbf {\bibinfo {volume}
  {60}},\ \bibinfo {pages} {905--916} (\bibinfo {year} {2016})}\BibitemShut
  {NoStop}%
\bibitem [{\citenamefont {Fall}\ \emph {et~al.}(2017)\citenamefont {Fall},
  \citenamefont {Lemaitre},\ and\ \citenamefont {Ovarlez}}]{FLO17}%
  \BibitemOpen
  \bibfield  {author} {\bibinfo {author} {\bibfnamefont {A.}~\bibnamefont
  {Fall}}, \bibinfo {author} {\bibfnamefont {A.}~\bibnamefont {Lemaitre}}, \
  and\ \bibinfo {author} {\bibfnamefont {G.}~\bibnamefont {Ovarlez}},\
  }\bibfield  {title} {\enquote {\bibinfo {title} {Discontinuous shear
  thickening in cornstarch suspensions},}\ }\href@noop {} {\bibfield  {journal}
  {\bibinfo  {journal} {EPJ Web of Conference: Powders and Grains 2017}\
  }\textbf {\bibinfo {volume} {140}},\ \bibinfo {pages} {09001} (\bibinfo
  {year} {2017})}\BibitemShut {NoStop}%
\bibitem [{\citenamefont {Royer}\ \emph {et~al.}(2016)\citenamefont {Royer},
  \citenamefont {Blair},\ and\ \citenamefont {Hudson}}]{RBH16}%
  \BibitemOpen
  \bibfield  {author} {\bibinfo {author} {\bibfnamefont {John~R.}\ \bibnamefont
  {Royer}}, \bibinfo {author} {\bibfnamefont {Daniel~L.}\ \bibnamefont
  {Blair}}, \ and\ \bibinfo {author} {\bibfnamefont {Steven~D.}\ \bibnamefont
  {Hudson}},\ }\bibfield  {title} {\enquote {\bibinfo {title} {Rheological
  signature of frictional interactions in shear thickening suspensions},}\
  }\href@noop {} {\bibfield  {journal} {\bibinfo  {journal} {Physical Review
  Letters}\ }\textbf {\bibinfo {volume} {116}} (\bibinfo {year}
  {2016})}\BibitemShut {NoStop}%
\bibitem [{\citenamefont {Weyand}\ \emph {et~al.}(2000)\citenamefont {Weyand},
  \citenamefont {Sternlight}, \citenamefont {Bellizzi},\ and\ \citenamefont
  {Wright}}]{WSBW00}%
  \BibitemOpen
  \bibfield  {author} {\bibinfo {author} {\bibfnamefont {Peter~G.}\
  \bibnamefont {Weyand}}, \bibinfo {author} {\bibfnamefont {Deborah~B.}\
  \bibnamefont {Sternlight}}, \bibinfo {author} {\bibfnamefont {Matthew~J.}\
  \bibnamefont {Bellizzi}}, \ and\ \bibinfo {author} {\bibfnamefont {Seth}\
  \bibnamefont {Wright}},\ }\bibfield  {title} {\enquote {\bibinfo {title}
  {Faster top running speeds are achieved with greater ground forces not more
  rapid leg movements},}\ }\href@noop {} {\bibfield  {journal} {\bibinfo
  {journal} {J. Applied Physiology}\ }\textbf {\bibinfo {volume} {89}},\
  \bibinfo {pages} {1991--1999} (\bibinfo {year} {2000})}\BibitemShut {NoStop}%
\end{thebibliography}

\end{document}